# Large multi-directional spin-to-charge conversion in low symmetry semimetal MoTe$_2$ at room temperature

C. K. Safeer[1,‡], Nerea Ontoso[1‡], Josep Ingla-Aynés[1], Franz Herling[1], Van Tuong Pham[1], Annika Kurzmann[2], Klaus Ensslin[2], Andrey Chuvilin[1,3], Iñigo Robredo[4,5], Maia G. Vergniory[3,4], Fernando de Juan[3,4], Luis E. Hueso[1,3,*], M. Reyes Calvo[1,3,6,*], Fèlix Casanova[1,3,*]

[1] CIC nanoGUNE, 20018 Donostia-San Sebastian, Basque Country, Spain.
[2] Solid State Physics Laboratory, ETH Zurich, 8093 Zurich, Switzerland
[3] IKERBASQUE, Basque Foundation for Science, 48013 Bilbao, Basque Country, Spain.
[4] Donostia International Physics Center (DIPC), 20018 Donostia-San Sebastian, Basque Country, Spain.
[5] Department of Condensed Matter Physics, University of the Basque Country (UPV/EHU), 48080 Bilbao, Basque Country, Spain.
[6] Departamento de Física Aplicada, Universidad de Alicante, 03690 Alicante, Spain

**Abstract:** Efficient and versatile spin-to-charge current conversion is crucial for the development of spintronic applications, which strongly rely on the ability to electrically generate and detect spin currents. In this context, the spin Hall effect has been widely studied in heavy metals with strong spin-orbit coupling. While the high crystal symmetry in these materials limits the conversion to the orthogonal configuration, unusual configurations are expected in low symmetry transition metal dichalcogenide semimetals, which could add flexibility to the electrical injection and detection of pure spin currents. Here, we report the observation of spin-to-charge conversion in MoTe$_2$ flakes, which are stacked in graphene lateral spin valves. We detect two distinct contributions arising from the conversion of two different spin orientations. In addition to the conventional conversion where the spin polarization is orthogonal to the charge current, we also detect a conversion where the spin polarization and the charge current are parallel. Both contributions, which could arise either from bulk spin Hall effect or surface Edelstein effect, show large efficiencies comparable to the best spin Hall metals and topological insulators. Our finding enables the simultaneous conversion of spin currents with any in-plane spin polarization in one single experimental configuration.

Symmetry is a unifying principle that governs all aspects of Physics, from the model of atomic orbitals to the Landau theory of phase transitions. The physical properties of crystalline solids are also highly constrained by symmetry[1] and, in turn, as symmetry is progressively lowered through the 32 crystallographic point groups, novel transport effects emerge, such as the non-linear Hall effect[2], the spin galvanic effect[3] and the valley magnetoelectricity[4] in non-centrosymmetric crystals, and the magnetochiral anisotropy[5] in chiral crystals. Crystal symmetry dictates also the geometry of a phenomenon that has attracted a lot of attention in recent years, the spin Hall effect (SHE) or its reciprocal effect [inverse SHE (ISHE)], which are generally observed in materials possessing strong spin-orbit coupling (SOC), and enables the interconversion between charge and spin currents. In conventional spin Hall materials, high crystal symmetry imposes that injecting a charge current density ($j_c$) can only result in a transverse spin current density ($j_s$) with a spin polarization ($s$) orthogonal to both $j_c$ and $j_s$ (Figure 1a)[6]. SHE/ISHE are crucial effects for the electrical generation or detection of spin current, required in applications such as spin-orbit torque memories[7,8,9] and spin-based logic devices[10,11]. These applications would highly benefit from more versatile SHE/ISHE configurations which can be obtained by lifting the constraints imposed by high crystal symmetry and enabling unusual spin-to-charge conversion geometries in low-symmetry crystals[12,13,14] (Figures 1b,c).

Transition metal dichalcogenides (TMDs), layered materials with strong SOC, are an ideal playground to observe not only conventional SHE, but also some of these unconventional SHE configurations [15–18] in their low-symmetry crystalline phases. One of such materials is MoTe$_2$, which appears in semimetallic, distorted 1T octahedral phases, namely 1T' and 1T$_d$ structures, which can be found at room and low temperatures, respectively. In its orthorhombic 1T$_d$ phase, MoTe$_2$ is an attractive material actively discussed in the context of Weyl physics[19,20]. A large spin Hall conductivity is theoretically predicted also for 1T$_d$-MoTe$_2$ [21], which has not been yet experimentally measured neither in this phase, nor in its 1T' monoclinic counterpart. Furthermore, in the 1T' phase with space group P2$_1$/m, which includes a screw axis along the Mo chain and a mirror plane perpendicular to this axis (Figures 1d,e)[22], certain unconventional SHE configurations are allowed (Figure 1b). In this Letter, we report the spin-to-charge conversion (SCC) of spin currents injected in MoTe$_2$ by spin absorption from graphene, with magnetic-field-induced full control of the spin polarization direction. Along with a highly efficient conventional SCC, we report the simultaneous observation of another unconventional SCC component (Figure 1c), unexpected as it remains forbidden by the mirror symmetry of the 1T' phase. The simultaneous detection of SCC originating from two different spin current polarization ($s$) directions for a fixed $j_c \perp j_s$ configuration, which can be only explained if we break the mirror plane of MoTe$_2$, opens new possibilities in the design of novel spintronic devices that could benefit from flexible geometries to generate, detect and control spin currents.

The spin absorption technique using lateral spin valves (LSVs) is a non-local method to study and quantify both the spin diffusion length and the SCC in strong SOC materials[23–25]. The advantage of using a non-local measurement is that spurious effects related to local currents, such as Oersted fields in spin-orbit torque techniques[26] or fringe-field-induced voltages in three-terminal potentiometric techniques[27,28] are avoided. Graphene-based LSVs have been employed to study prototypical spin Hall metals as Pt[29,30] due to the excellent spin transport properties of graphene[31–35], and they are expected to be ideal for studying TMDs as the van der Waals interaction with graphene[36–41] allows the formation of a good contact between them. Using this technique, both spin-to-charge and charge-to-spin conversions can be studied by exchanging the spin injection and detection terminals. For the sake of simplicity, we will focus on spin-to-charge conversion measurements, in a scheme sketched in Figures 1g-j. An electrical current applied from a ferromagnetic electrode (in our case, Co) into the graphene channel creates a spin accumulation at the interface, which diffuses away as $j_s$ in the graphene channel. $j_s$ is then absorbed into the SOC material (MoTe$_2$) along $z$ and converted into $j_c$, which is detected as a non-local voltage ($V_{NL}$) across the MoTe$_2$ flake along $y$. By applying a magnetic field along the different directions ($x$, $y$, and $z$), we control the direction of the spin polarization $s = (s_x, s_y, s_z)$ of the spin current absorbed into MoTe$_2$ (Figures 1g-j). On the one hand, due to the shape anisotropy, the easy axis of the Co electrode, and thus its magnetization at zero field, lies along the $y$-axis. By applying enough magnetic field along $\pm y$-direction ($B_y$), the magnetization of the Co electrode and thus $s = (0, \pm s_y, 0)$ of the injected spins can be switched (the coercive field of Co electrodes is typically < 0.05 T). However, $B_y$ does not induce spin precession during the spin transport along the graphene channel because $s$ and $B_y$ are parallel to each other and, as a consequence, $j_s$ absorbed into the MoTe$_2$ has one of the two polarization directions ($\pm s_y$) (Figure 1g). On the other hand, applying a magnetic field either along the in-plane $x$-axis ($B_x$) or along the out-of-plane $z$-axis ($B_z$), i.e. hard axes of the Co electrode, affects the polarization orientation of the injected spins in two different ways. At zero magnetic field, $s$ lies along the Co easy axis, then, an applied field $B_x$ or $B_z$ can change $s$ by 1) inducing spin precession in the graphene channel along the $y - z$ plane, as shown in Figure 1h, or along the $x - y$ plane, as shown in Figure 1j, respectively, or by 2) changing the magnetization of the Co electrode and thus the polarization of the injected spins along the $x$ or $z$ directions, respectively (see Figure 1i for $x$-direction). The final $s$ of the spin current absorbed into the MoTe$_2$ is set by the combination of these two processes. Note that as the magnetization of the Co (and thus the injected spins) starts to align with the field direction, $s_y$ decays. Hence the precession disappears at the saturation magnetic field, when $s$ and the field become parallel. The strength of the field required to change the magnetization along the hard axis depends on the shape anisotropy of the Co electrode along that direction[42]. In our case, a relatively small field is required (>0.3 T) to saturate the Co magnetization along the $x$-axis compared to that along the $z$-axis (>1.5 T). Therefore, in the low $B_x$ regime, the overall $s$ is determined by the variation of the magnetization of the Co injector and the spin precession (Figures 1h,i), whereas at the low $B_z$ regime, the direction of $s$ is set by the spin precession only (Figure 1j).

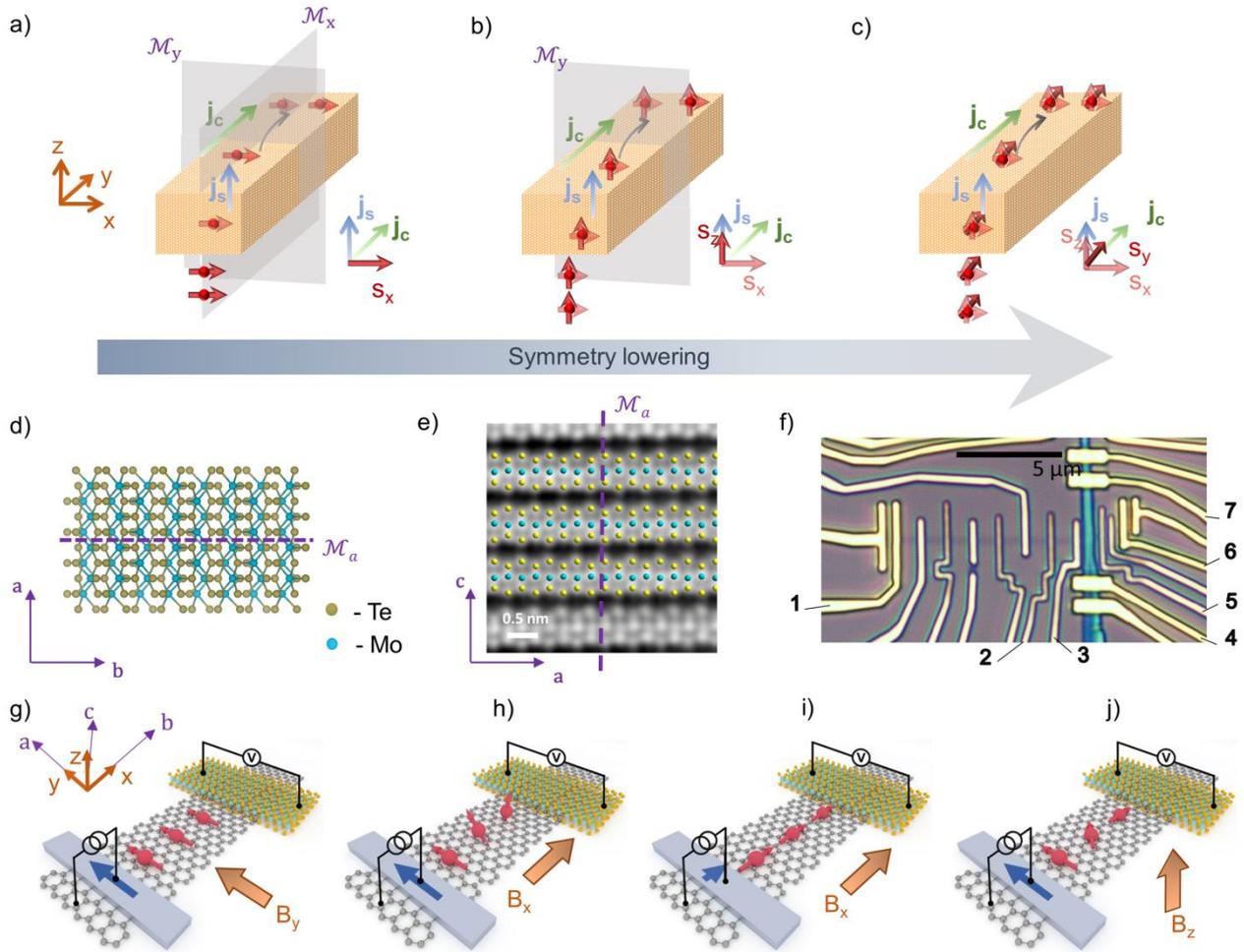

**Figure 1.** (a-c) Sketch of spin Hall effects when lowering the crystal symmetry. (a) In high symmetry crystals with at least two mirror planes of symmetry, $M_x$ and $M_y$, only the conventional SHE is allowed: a charge current density ($j_c$) applied along $y$-axis results in the out-of-plane spin current density ($j_s$) with spin polarization ($s$) along $x$. (b) By reducing the crystal symmetry to one single mirror plane, new configurations are permitted; for example, $j_c$ perpendicular to the mirror plane $M_y$ can result in $j_s$ parallel to $s$. On the contrary, and in this same example, symmetry upon reflection prevents the generation of $j_s$ with $s$ to point parallel to $j_c$. (c) This is, however, also allowed if the remaining mirror symmetry is broken, and spin-to-charge current conversion becomes now possible for all three spin orientations. (d) The $a - b$ plane of the monoclinic 1T´ crystal structure of MoTe$_2$. The only mirror plane of this structure ($M_a$) is also shown. (e) TEM image of one of our exfoliated MoTe$_2$ flakes, with the $a - c$ plane of the 1T´ crystal structure superimposed. (f) Optical microscope image of one of our LSV devices (device 1), where the spin absorption technique is performed. It contains a graphene channel with a MoTe$_2$ flake (green) placed on top. The ends of the graphene channel and MoTe$_2$ are connected to Ti/Au contacts (yellow). Several Co/TiO$_x$ electrodes are placed on top of the graphene channel (white). (g-j)

Sketches of the different measurement configurations that result in spin-to-charge conversion in our device. The $a$ and $b$ axes of the MoTe$_2$ crystal lie along the $y$ and $x$ axes of the experimental configuration.

An optical microscope image of one of our devices (details of the device fabrication are explained in Methods) is shown in Figure 1f. An exfoliated few-layer graphene flake (590-nm-wide) works as the spin transport channel. A needle-like flake of MoTe$_2$ (890-nm-wide and 11-nm-thick) that acts as the SOC material is placed on top of the graphene flake, forming a van der Waals heterostructure. Several Co electrodes that work as spin injectors/detectors are also placed on top of graphene. From transmission electron microscopy (TEM) (Figure 1e) and polarized Raman spectroscopy, we confirm that the long direction of the needle-shaped flake corresponds to the $a$-axis of the MoTe$_2$ crystal, i.e., the Mo zig-zag chain is aligned along $y$, parallel to the easy axis of the Co electrode (see Note S1). The graphene channel and MoTe$_2$ are contacted with Ti/Au electrodes. At first, by comparing the spin transport between the Co electrodes with (Co-3 and Co-5 in Figure 1f) and without (Co-2 and Co-3) the MoTe$_2$ flake placed in between, we confirm that most of the spin current gets absorbed (along $z$) into the MoTe$_2$ (see Note S4). From this comparison, we should extract the spin diffusion length of MoTe$_2$ ($\lambda_s^{MoTe_2}$). Unfortunately, the graphene/MoTe$_2$ interface resistance is not negligible (see Note S2 for the proper calculation of this quantity) and dominates the spin absorption, thus preventing us from extracting $\lambda_s^{MoTe_2}$. Nevertheless, since $\boldsymbol{j_s}$ is absorbed into the MoTe$_2$, SCC is expected to occur, as discussed in detail in Note S13. To perform such measurement, we apply a constant current $I = 10$ μA between the Co injector (Co-3) and graphene (Ti/Au contact 1) while we measure the voltage drop $V_{NL}$ proportional to the generated charge current $\boldsymbol{j_c}$ across the MoTe$_2$ (Ti/Au contacts 4 and 7). The SCC signal is represented by a non-local resistance ($R_{NL} = V_{NL}/I$). In all our measurements, the directions of $\boldsymbol{j_s}$ (along $z$) and the measured $V_{NL}$ (along $y$) are fixed, whereas the orientation of $\boldsymbol{s}$ is controlled with the magnetic field applied along the three directions ($x$, $y$, and $z$) as discussed above.

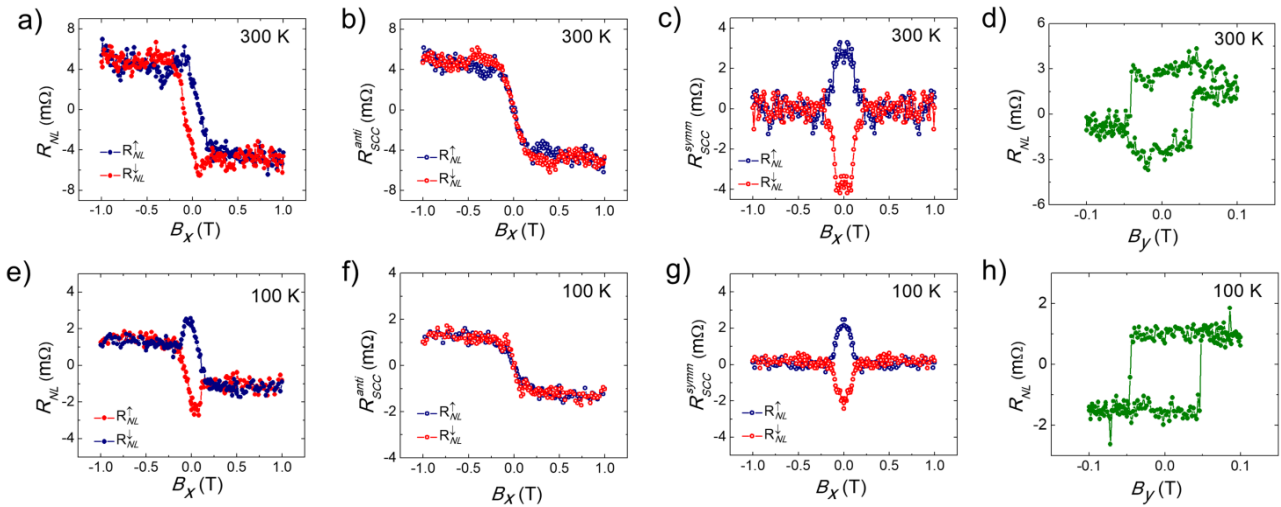

**Figure 2.** (a) Nonlocal spin-to-charge conversion resistance ($R_{NL}$) measured at room temperature as a function of the magnetic field applied along $x$ ($B_x$), i.e., the in-plane hard axis of the Co electrode, for initial magnetization of the Co electrode saturated along positive (blue) and negative (red) $y$-direction. The measurement configuration is shown in Figures 1h,i. A baseline of −80 mΩ is subtracted for both curves. (b) Antisymmetric components ($R_{SCC}^{anti}$) of the two curves shown in panel a. (c) Symmetric components ($R_{SCC}^{symm}$) of the two curves shown in panel a. (d) Nonlocal spin-to-charge conversion resistance ($R_{NL}$) measured at room temperature as a function of the magnetic field applied along $y$ ($B_y$), i.e., the easy axis of the Co electrode. The measurement configuration is shown in Figure 1g. A baseline of −36 mΩ is subtracted. (e) The same $R_{NL}$ vs. $B_x$ curves described in panel a measured at 100 K. A baseline of −42 mΩ is subtracted for both curves. (f)

Antisymmetric components of the two curves shown in panel e. (g) Symmetric components of the two curves shown in panel e. (h) The same $R_{NL}$ vs. $B_y$ curve described in panel d measured at 100 K. A baseline of −15 mΩ is subtracted.

First, we study the SCC using the conventional experimental geometry[29,30], *i.e.* by sweeping $B_x$, and the resulting curves $R_{NL}$ vs. $B_x$ for two different temperatures are shown in Figures 2a (300 K) and 2e (100 K). The curves at other temperatures are shown in Note S5. The measurement was performed in 4 steps: For the first two steps, $R_{NL}$ was measured while sweeping $B_x$ from 0 to 1 T and 0 to −1 T, with the magnetization of Co initially saturated in the +$y$-direction prior to each sweep ($R_{NL}^{\uparrow}$, blue curve). Then similar measurements were repeated for the magnetization of the Co initially set along the −$y$-direction ($R_{NL}^{\downarrow}$, red curve). According to the conventional geometry of SCC ($j_s$, $j_c$ and $s$ must be mutually orthogonal), in our experiment the voltage $V_{NL}$ measured along $y$ with $j_s$ fixed along $z$ is expected to detect the $x$-component of $s$ ($s_x$). As depicted in Figure 1i, $B_x$ causes the Co magnetization to rotate towards $x$ and saturates when $|B_x| > 0.3$ T (see Note S3 for the experimental determination of the angle $\theta_M$ between the easy axis and the Co magnetization as a function of $B_x$). The corresponding variation of $s_x$ is proportional to $\sin \theta_M$ and expected to result in an $R_{NL}$ vs. $B_x$ curve that varies linearly below the saturation field, crosses zero for $B_x = 0$ and saturates above the saturation field[29,30]. Figures 2a,e show indeed the saturation of $R_{NL}$ when $|B_x| > 0.3$ T. However, below that saturation value, even at $B_x=0$, we observe a non-zero value of $R_{NL}$ that evolves differently with increasing $B_x$ and depends on the initial Co magnetization aligned along +$y$ and −$y$ (red and blue curve, respectively, in Figure 2), suggesting the existence of a non-zero SCC signal arising from the $y$-component of $s$ ($s_y$) as well. To verify this possibility, we extract the antisymmetric (Figures 2b,f) and symmetric (Figures 2c,g) components of the $R_{NL}$ vs. $B_x$ curves. The first contribution, $R_{SCC}^{anti}$, shows the expected behavior proportional to $s_x$ and it is independent of the initial magnetization direction, indicating SCC in MoTe$_2$ with the conventional symmetry. However, the second contribution, $R_{SCC}^{symm}$, shows a behavior which depends on the initial magnetization direction: $R_{NL}^{\uparrow}$ is maximum at $B_x = 0$ and decreases to zero at saturation, whereas $R_{NL}^{\downarrow}$ shows the same shape with opposite sign. They are thus proportional to the variation of $s_y$, indicating a new and unconventional SCC in our system with a $j_c \| s$ geometry. To further confirm the SCC in this unconventional configuration, we measure $R_{NL}$ while sweeping $B_y$ (see sketch in Figure 1g). Very interestingly, we observe a square hysteresis loop for $R_{NL}$ vs. $B_y$, shown in Figures 2d (300 K) and 2h (100 K), with the signal switching sign at the coercive field of the Co electrode (~0.034 T). The curves at other temperatures are shown in Note S6. We also note an opposite sign between the measured $V_{NL}$ corresponding to the SCC with $s_x$ (Figures 2b,f) and $s_y$ (Figures 2d,h), indicating that the sign of the conventional and unconventional SCCs are opposite. The same measurements were performed in another device (device 2) and obtained qualitative the same results (see Note S9). We also performed two separate SCC measurements by injecting spins from two different Co electrodes with different coercivity and detected at the same MoTe$_2$ flake. We observe that the switching fields for these two $R_{NL}$ vs. $B_y$ loops follow the coercivity of each Co injector (see Note S10). This set of control experiments shows the robustness of this unconventional SCC signal in our system. By exchanging the current and voltage terminals, we measured both the direct (charge-to-spin) and the inverse (spin-to-charge) conversion for both the conventional and unconventional SCC components, which fulfills the expected Onsager reciprocity (see Note S11).

In the $R_{NL}$ measurements presented above, $s$ is set by the variation of magnetization of the Co injector. In the $R_{NL}$ vs. $B_x$ curves, the variation of $R_{NL}$ associated to SCC from $s_x$ ($R_{SCC}^{anti}$, Figure 2b) is proportional to the variation of the magnetization of the Co injector along the $x$-direction. However, the decrease of $R_{SCC}^{symm}$ from the maximum to zero with $B_x$ (Figure 2c) can be indistinguishably associated to the decrease of $s_y$ either via the decrease of the Co magnetization along the $y$-direction or the spin precession in the $y-z$ plane. It is thus important to verify that the measured $R_{NL}$ signal corresponds to SCC and it is not originated from an artifact related to the Co magnetization, such as the anomalous Hall effect or the anisotropic magnetoresistance. A measurement sensitive only to the spin precession is required to confirm that the true origin of the two observed contributions is SCC. It is thus convenient to perform a measurement by applying $B_z$. The resulting curves $R_{NL}$ vs. $B_z$ at two different temperatures are shown in Figures 3a (300 K) and 3d (100 K). The curves at other temperatures are shown in Note S7. The measurement was again performed in 4 steps: for the first two steps, $R_{NL}$ was measured while sweeping $B_z$ from 0 to 2 T and 0 to $-2$ T, initially saturating the Co magnetization along $+y$-direction prior to each sweep ($R_{NL}^\uparrow$, blue curve). Then the protocol was repeated for the Co magnetization initially set along $-y$-direction ($R_{NL}^\downarrow$, red curve). At low $B_z$ regime, the direction of $\boldsymbol{s}$ is dictated by the spin precession along the $x-y$ plane and the variation of the magnetization of Co along the $z$-direction can be neglected because the saturation occurs at much larger values (>1.5 T). To understand the overall $R_{NL}$ signal (Figures 3a,d), we need to consider that the two SCC components (arising from $s_x$ and $s_y$) contribute to $R_{NL}$, but each contribution has a different dependence on increasing $B_z$. In particular, the contribution from $s_y$ to SCC should give rise to a symmetric Hanle precession curve (analogous to the standard Hanle curves obtained in a conventional LSV, where the Co detector is also sensitive to $s_y$, see Note S3), whereas the contribution from $s_x$ should give rise to an antisymmetric Hanle precession curve (as expected when the detector is sensitive to an $\boldsymbol{s}$ component which is perpendicular to the injected $\boldsymbol{s}$)[30,40]. Both curves should also change sign when the initial magnetization direction is switched because the injected spins are opposite. This allows us to remove the background by subtracting the two curves and obtain the pure SCC signal, $R_{SCC} = (R_{NL}^\uparrow - R_{NL}^\downarrow)/2$ (Figures 3b,f). Then, symmetrization and antisymmetrization of the $R_{SCC}$ curve allow us to distinguish the SCC contributions due to $s_x$ (Figures 3c,g) and $s_y$ (Figures 3d,h), which have the expected corresponding behavior as a function of $B_z$. From this result, we confirm that the measured $V_{NL}$ signal indeed corresponds to the sum of a conventional $[(\boldsymbol{j_s} \parallel z) \perp (\boldsymbol{j_c} \parallel y) \perp (\boldsymbol{s} \parallel x)]$ and an unconventional $[(\boldsymbol{j_s} \parallel z) \perp (\boldsymbol{j_c} \parallel \boldsymbol{s} \parallel y)]$ SCC component, ruling out any artifacts not coming from spin currents. As observed before, we confirm that $R_{SCC}^{symm}$ and $R_{SCC}^{anti}$ have opposite sign (see Note S15 for details). A misorientation between the main crystallographic axis of the MoTe$_2$ needle-like flake with respect to the magnetic easy axis of Co injector, in combination with the anisotropy between the conventional (orthogonal) spin Hall conductivity tensor elements in 1T'–MoTe$_2$[21] could in principle lead to the observed unconventional SCC. However, the alignment accuracy of <1° with the fabrication of the Co electrodes (determined from the optical microscopy image) rules out this possibility. The possibility of a tilt in the magnetic easy axis of Co injector towards the hard axis ($x$-axis) to explain the observed result is also discarded (See Note S3 for details).

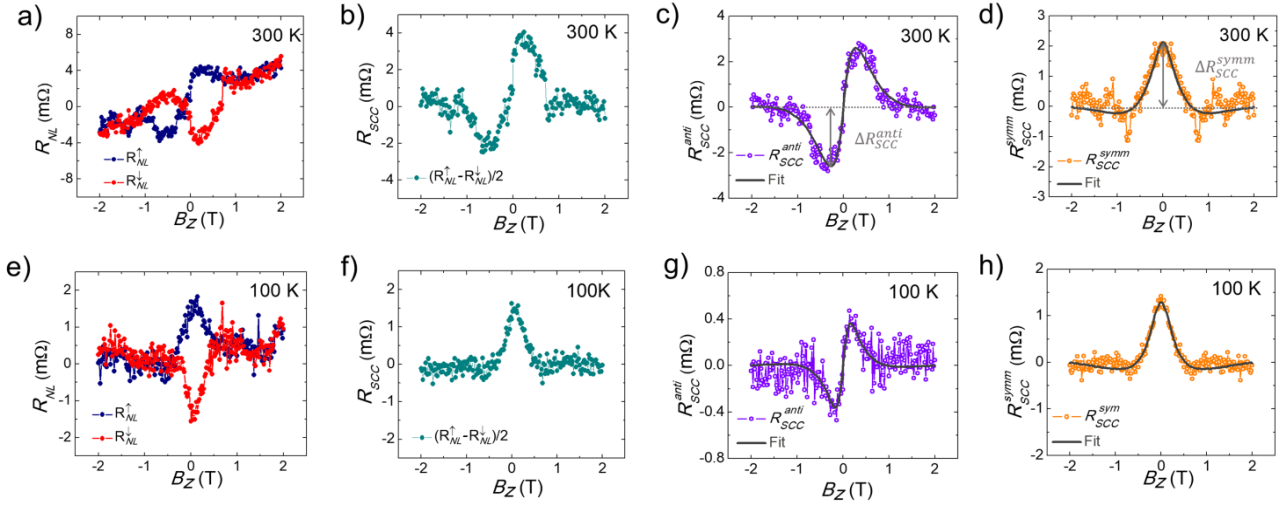

**Figure 3.** (a) Nonlocal spin-to-charge conversion resistance ($R_{NL}$) measured at room temperature as a function of the magnetic field applied along the $z$-direction ($B_z$), *i.e.*, the out-of-plane hard axis of the Co electrode, for initial positive (blue) and negative (red) magnetization directions of the Co electrode. The measurement configuration is shown in Figure 1j. A baseline of −40 mΩ is subtracted for both curves. (b) The SCC signal $R_{SCC} = (R_{NL}^\uparrow - R_{NL}^\downarrow)/2$ obtained from the data in panel a. (c) Antisymmetric component of the $R_{SCC}$ curve shown in panel b, which is fitted with the solution of the Bloch equation (black solid curve). It corresponds to the contribution of $s_x$ to SCC. The amplitude of the signal $\Delta R_{SCC}^{anti}$, defined from the background to the peak at negative $B_z$ because this sets the spin polarization along $+x$-direction, is negative. (d) Symmetric component of the $R_{SCC}$ curve shown in panel b, which is fitted with the solution of the Bloch equation (black solid curve). It corresponds to the contribution of $s_y$ to SCC. The amplitude of the signal $\Delta R_{SCC}^{symm}$, defined from the background to the peak at zero field, is positive. (e) $R_{NL}$ vs. $B_z$ curves described in panel a measured at 100 K. A baseline of −15.5 mΩ is subtracted for both curves. (f) The SCC signal $R_{SCC} = (R_{NL}^\uparrow - R_{NL}^\downarrow)/2$ curve obtained from the data in panel e. (g) Antisymmetric component of the $R_{SCC}$ curve shown in panel f, which is fitted with the solution of the Bloch equation (black solid curve). (h) Symmetric component of the $R_{SCC}$ curve shown in panel f, which is fitted with the solution of the Bloch equation (black solid curve).

In our experimental geometry, both the conventional and unconventional SCCs can originate from two different mechanisms, the spin Hall effect at the bulk and the Edelstein effect at the surface states. An unconventional charge-to-spin conversion effect has been reported in the low-symmetry 1T$_d$ phase of WTe$_2$[15,16] and 1T´ phase of MoTe$_2$[43] using spin-orbit torque measurements, where out-of-plane damping-like torque, in which the spin current is parallel to the out-of-plane spin polarization ($j_s \| s \| z$), is observed. Such torque is allowed in crystals with a single mirror plane. This component is also allowed in our measurement geometry, since $j_c$ is measured along the $a$-axis, *i.e.*, across the only mirror plane $M_a$ of the MoTe$_2$ flake (see Figure 1b). It is interesting to note that this contribution of $s_z$ to SCC, which should appear as an antisymmetric Hanle precession in the $R_{NL}$ vs. $B_x$ curve, is not observed in our experiments (see Figures 2b,f). Taking into account that this SCC component is reported to be ~8 times smaller than the conventional one[43], the corresponding signal is likely below the resolution of our measurement. In contrast, in the observed unconventional SCC, the charge current is parallel to the in-plane spin polarization ($j_c \| s \| y$) a case which is forbidden by symmetry both in the bulk and the surface states of 1T´–MoTe$_2$ (see Note S16). A recent work reports an unexpected SCC component similar to our case ($j_c \| s$) in 1T$_d$–WTe$_2$ below 100 K when applying $j_c$ parallel to the mirror plane and attributes it to the special spin texture of the topological Fermi arcs in the low-temperature Weyl semimetal phase[17]. However, this does not apply to our case. Although a transition into a Weyl semimetal phase associated with a structural transition from 1T´ to 1T$_d$ below 250 K has been reported in MoTe$_2$[19,20], our measurements are performed in the 1T´ phase (no phase transition to

$1T_d$ is observed in our flake, see Note S2, in agreement with Ref.[43]), where no topological surface Fermi arcs are expected. Nevertheless, to check whether the trivial surface states (via inverse Edelstein effect) could still give rise to the unconventional SCC observed in MoTe$_2$, we computed the band structure and spin polarizations for 1T´–MoTe$_2$ (see Note S16). Our calculation confirms that it is not possible to have SCC at the surface states with a ($\boldsymbol{j_c} \parallel \boldsymbol{s} \parallel y$) configuration when $\boldsymbol{j_c}$ is applied along the $y$-axis (perpendicular to the mirror plane). Indeed, this conclusion applies for any structure with mirror symmetry, as it is the case of the bulk and surfaces of 1T' and 1T$_d$ phases. Therefore, in order to explain our observation, the mirror symmetry must be broken in some way. One possibility is that the MoTe$_2$ flake has developed shear strain[18] from the mismatch with the graphene lattice or during the stamping process to the substrate. In order to check whether such strain would allow the unconventional SCC component, we repeated the previous calculations by including now shear strain (see Note S16). We confirm that spin polarizations do not show any symmetry beyond time reversal, and $\boldsymbol{j_c}$ applied along the $y$-direction can generate finite averages of all components, including $s_y$. Since the only remaining symmetry after shear strain is inversion symmetry, all possible SCC components are also allowed to originate from the bulk states via the spin Hall effect.

Figure 4a shows the temperature dependence, from 300 K down to 75 K, of the amplitude $\Delta R_{SCC}$ of the two observed SCC components (as defined in Figures 3c,d). For both cases, we observe that the SCC signal increases with temperature, reaching the largest value at room temperature. A comparison of the amplitudes extracted from all measurement configurations is shown in Note S8. Such trend arises not only from the temperature dependence of the conversion efficiency, but also of the spin transport parameters of Co and graphene, the resistivity of MoTe$_2$ and the interface resistance at the graphene/MoTe$_2$ van der Waals gap. The spin transport parameters were calculated by fitting the nonlocal Hanle spin precession measurement across the reference LSVs (Co-2 and Co-3) with the solutions of Bloch equations (see Note S3). We obtained a spin lifetime of $\tau_s^{gr}$~100 ps and a spin diffusion constant of $D_s^{gr}$~1.5× $10^{-3}$ m$^2$/s for the graphene and a spin polarization of $|P|$~4.15% for the Co/graphene interface at room temperature (see Note S3 for the other temperatures). The different resistances were obtained with 4-point electrical measurements and, for the interface resistance, we performed a finite element 3D modeling (see Note S2). Since we cannot, a priori, distinguish whether the SCCs occur at the bulk (via ISHE) or at the surface (via inverse Edelstein effect), we analyze our data assuming bulk ISHE of MoTe$_2$ to extract the efficiency given by the spin Hall angle ($\theta_{ij}^k$, where $i, j, k$ are the directions of $\boldsymbol{j_s}, \boldsymbol{j_c}$ and $\boldsymbol{s}$, respectively). In this case, we fit the antisymmetric (Figures 3c,g) and symmetric (Figures 3d,f) components of the SCC precession curves to the solution of the Bloch equation to extract $\theta_{zy}^x$ and $\theta_{zy}^y$, respectively (Note S14). These values will depend on the value of $\lambda_s^{MoTe_2}$, which is not possible to determine from spin absorption, as discussed Notes S4 and S13. Moreover, the absolute sign of $\theta_{ij}^k$ cannot be determined because the sign of $P$ is not known. However, we can extract $|\theta_{ij}^k|$ for a broad range of $\lambda_s^{MoTe_2}$ values (Note S14). On the one hand, we find that, for $\lambda_{MoTe_2}$ values similar or longer than the thickness of MoTe$_2$ (11 nm), $|\theta_{ij}^k|$ tends to a low constant value independent of $\lambda_s^{MoTe_2}$, because the SCC process is limited by the MoTe$_2$ thickness. This behavior allows us to get a lower limit for $|\theta_{ij}^k|$, which for the case of the conventional component is $|\theta_{zy}^x| \geq 0.21$ at room temperature, comparable to the best known spin Hall metals[24,25] and alloys[23,44]. The lower limit for the unconventional component efficiency is also found to be remarkably large ($|\theta_{zy}^y| \geq 0.10$), and with opposite sign, at room temperature. The opposite sign between the conventional and unconventional SCCs observed in our case is not surprising. Indeed, theoretical calculations in MoTe$_2$ show that even the sign of the conventional spin Hall conductivities with mutually orthogonal symmetries along different crystal axes can be opposite[21]. On the other hand, for $\lambda_s^{MoTe_2}$ values much smaller than the MoTe$_2$ thickness, it is the $|\theta_{ij}^k|\lambda_s^{MoTe_2}$ product that tends to a low constant value, which is what defines the SCC efficiency[45,46]. This allows us to give a lower limit for $|\theta_{ij}^k|\lambda_s^{MoTe_2}$, which for the case of the conventional component is $|\theta_{zy}^x|\lambda_s^{MoTe_2} \geq 1.15$ nm at room temperature, much larger than heavy metals (~0.1–0.3 nm)[24,25]. If we assume that the SCC originates at the surface states of MoTe$_2$, this product would correspond to the Edelstein length ($\lambda_{IEE}$), the efficiency associated to the inverse Edelstein effect. This value would be comparable to the best efficiency reported in topological insulators ($\lambda_{IEE} = 2.1$ nm for α–Sn[47]). The lower limit for the unconventional component efficiency is $|\theta_{zy}^y|\lambda_s^{MoTe_2} \geq 0.5$ nm at room temperature (with opposite sign to the conventional one). Regardless of the origin of the SCC (bulk or surface), this quantification demonstrates that MoTe$_2$ is a very promising material for spintronics applications with the flexibility of obtaining very large SCCs for spins with any in-plane spin polarization at room temperature.

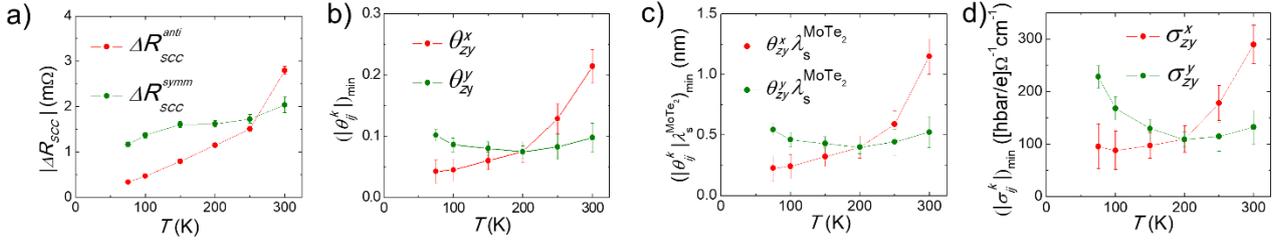

**Figure 4.** Temperature dependence of: (a) the amplitude $\Delta R_{SCC}$ of the two components of the SCC signal, obtained from the antisymmetric and symmetric components of the $R_{SCC}$ vs $B_z$ curves as the ones shown in Figure 3c and 3d, respectively; (b) lower limit of the spin Hall angle of MoTe$_2$ ($\theta_{ij}^k$); (c) lower limit of the product of the spin Hall angle and the spin diffusion length of MoTe$_2$ ($\theta_{ij}^k\lambda_s^{MoTe_2}$); (d) lower limit of the spin Hall conductivity ($\sigma_{ij}^k$). All plots show the result for the $s_x$ component (red curves) and $s_y$ component (green curves) of the SCC, which have opposite sign to each other.

By performing the fittings of the symmetric and the antisymmetric part of the SCC precession curves ($R_{SCC}$ vs. $B_z$) at different temperatures between 300 K and 75 K, the lower limits of $|\theta_{ij}^k|$ and $|\theta_{ij}^k|\lambda_s^{MoTe_2}$ are calculated and plotted in Figure 4b and 4c, respectively. The trend of the efficiency (either $|\theta_{ij}^k|$ or $|\theta_{ij}^k|\lambda_s^{MoTe_2}$) is different for the two different SCC components: whereas the unconventional one does not vary much with temperature, the standard one increases with temperature. A better parameter to characterize the bulk SHE is the spin Hall conductivity ($\sigma_{ij}^k = \theta_{ij}^k/\rho_{jj}$, where $\rho_{jj}$ is the longitudinal resistivity along $j$), which is plotted in Figure 4d. In metals, $\sigma_{ij}^k$ is expected to be temperature independent when the intrinsic contribution arising from the Berry curvature of the band structure dominates[24,25]. This does not seem to be the case for any of the components in MoTe$_2$: whereas the unconventional component ($|\sigma_{zy}^y|$) decreases with temperature, the conventional one ($|\sigma_{zy}^x|$) increases. These trends can be explained when an additional mechanism contributes to the SCC. One possibility is an extrinsic (skew scattering and/or side jump) contribution to the SHE with opposite sign[25,48]. Another possibility is the presence of the inverse Edelstein effect in graphene induced by spin-orbit proximity with MoTe$_2$, which would be detected with the same measurement configuration and has been recently reported in other graphene/TMDs van der Waals heterostructures[40,49–51]. This contribution decreases with increasing temperature[40,51] and, therefore, an opposite (same) sign with respect to the intrinsic $\sigma_{zy}^x$ ($\sigma_{zy}^y$) could explain the observed trend. Finally, it is also possible that the intrinsic spin Hall conductivity components of semimetal MoTe$_2$ have a stronger temperature dependence than that of metals. Further work would be required for a deeper understanding of the different mechanisms contributing to the SCC in MoTe$_2$.

In conclusion, we observe and quantify spin-to-charge conversion in MoTe$_2$. Along with a large efficiency in the conventional orthogonal configuration (spin current, charge current and spin polarization are mutually perpendicular), we also demonstrate SCC in an unusual non-orthogonal geometry, with a charge current arising parallel to the spin orientation. Whereas the low crystal symmetry of MoTe$_2$ allows for non-orthogonal SCC configurations, the unconventional SCC observed here is only possible if even the remaining mirror crystal symmetry of MoTe$_2$ is broken in our samples, likely due to fabrication-induced shear strain. Regardless of the origin, we present here a system where any in-plane polarization of injected spins results into charge conversion, bringing new flexibility to the design of spin logic devices. Inversely, the ability to obtain spin currents with any in-plane spin polarization by applying electrical current along a single direction is a promising feature for spin-orbit torque memories, current-induced domain wall and skyrmions motion-related applications. All the above makes MoTe$_2$ a system of interest for further investigation and a promising material for future spintronics applications.

**Methods.** *Device fabrication.* The graphene/MoTe$_2$ van der Waals heterostructures are fabricated by mechanical exfoliation followed by dry viscoelastic stamping inside a glove box with inert Ar atmosphere. We first exfoliate graphene from bulk graphitic crystals (supplied by NGS Naturgraphit GmbH) using a Nitto tape (Nitto SPV 224P) onto Si substrates with 300 nm SiO$_2$. Few-layer graphene flakes are identified by optical contrast under an optical microscope. Then a MoTe$_2$ crystal (supplied by HQ Graphene) is exfoliated using the Nitto tape and transferred on to a piece of poly-dimethyl siloxane (Gelpak PF GEL film WF 4, 17 mil.). After identifying a proper MoTe$_2$ flake with a short and narrow part using an optical microscope, it is stamped on top of graphene using visco-elastic stamping tool where a three-axes micrometer stage is used to accurately position the two flakes. The graphene is then connected with Ti(5nm)/Au(100nm) contacts fabricated using electron-beam lithography followed by electron beam deposition in ultrahigh vacuum and lift-off. Using the same procedure, the 35 nm thick Co electrodes are fabricated on top of the graphene channel. Before this deposition, a TiO$_x$ tunnel barrier is fabricated by depositing 3Å of Ti and subsequent natural oxidation in air. The widths of the Co electrodes vary between 250 nm to 400 nm leading to different coercive fields for each electrode. The exact dimensions of the devices are extracted from atomic force microscopy after they have been measured (Note S1).

*Materials characterization.* **Raman characterization**: The characterization of the crystallographic axis of the MoTe$_2$ flakes are performed using a WITec confocal microscope with a green polarized laser (532nm) and placing a linear polarizer before the spectrograph, parallel to the laser polarization. By rotating the sample, we analyze the Raman spectrum of the flake for different angles between the polarized light and the crystallographic axis and thus, check the crystallographic axis along which the material cleaves (Note S1). **STEM Characterization:** STEM study has been performed with Titan 60-300 TEM/STEM instrument (FEI, Netherlands) in STEM mode with an acceleration voltage of 300kV. The cross-sectional sample of MoTe$_2$ stripe on SiO$_2$ has been cut by FIB strictly along the cleaved edge in order to determine the crystal orientation of the flake (Note S1).

*Electrical measurements.* Charge and spin transport measurements are performed in a Physical Property Measurement System by Quantum Design, using a 'DC reversal' technique with a Keithley 2182 nanovoltmeter and a 6221 current source at temperatures ranging from 75 K to 300 K. We apply in-plane and out-of-plane magnetic fields with a superconducting solenoid magnet and a rotatable sample stage.

- **ASSOCIATED CONTENT**

**Supporting Information**

Sample characterization including AFM and STEM imaging, Raman spectroscopy; temperature dependence of sheet resistance of graphene, resistivity of MoTe$_2$ and graphene MoTe$_2$ interface resistance; spin transport in the reference graphene channel; spin absorption in MoTe$_2$; temperature dependence of $R_{NL}$ vs. $B_x$ measurements; temperature dependence of $R_{NL}$ vs. $B_y$ measurements; temperature dependence of $R_{NL}$ vs. $B_z$ measurements; comparison of amplitudes of the SCC signal of $s_y$ and $s_x$ obtained from $R_{NL}$ vs. $B_x$, $R_{NL}$ vs. $B_y$ and $R_{NL}$ vs. $B_z$ measurements; reproducibility showing experiments in device 2; $R_{NL}$ vs. $B_y$ measurement by injecting spin current from two Co electrodes with different coercive fields; Onsager reciprocity; model for spin-to-charge conversion in MoTe$_2$ with an interfacial barrier between the graphene and MoTe$_2$; In-plane angle dependence of R$_{NL}$; Density functional theory (DFT) calculations.

- **AUTHOR INFORMATION**

**Corresponding Authors**
*E-mail: l.hueso@nanogune.eu
*E-mail: r.calvo@nanogune.eu
*E-mail: f.casanova@nanogune.eu

**Author Contributions**
‡These authors contributed equally to this work.

**Notes**
The authors declare no competing financial interest.

- **ACKNOWLEDGMENTS**

The authors thank Roger Llopis for drawing the sketches used in the figures. This work is supported by the Spanish MINECO under the Maria de Maeztu Units of Excellence Programme (MDM-2016-0618) and under Projects MAT2015-65159-R, RTI2018-094861-B-100, MAT2017-88377-C2-2-R and MAT2017-82071-ERC, and by the European Union H2020 under the Marie Curie Actions (794982-2DSTOP and 766025-QuESTech). N.O. thanks the Spanish MINECO for a Ph.D. fellowship (Grant No. BES-2017-07963). V.T.P. acknowledges postdoctoral fellowship support "Juan de la Cierva-incorporación" program by the Spanish MINECO (Grant No. FJCI-2017-34494).

- **REFERENCES**

# SUPPORTING INFORMATION

# S1. Sample characterization

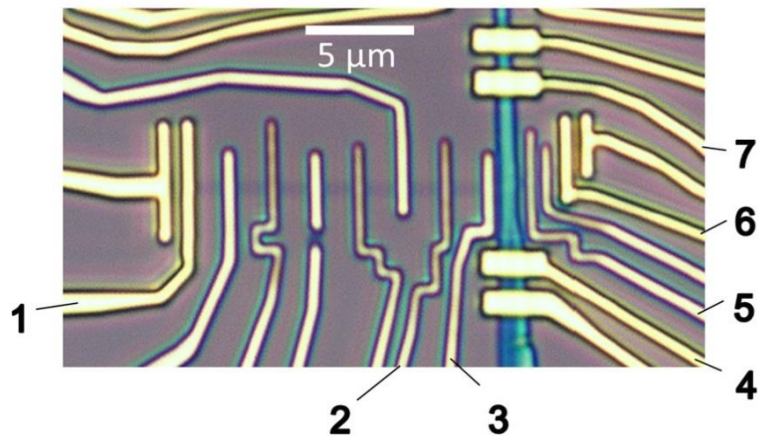

**Figure S1. Optical image of device 1.** The LSV with Co electrodes 2 and 3 is used as a reference. The LSV with Co electrodes 3 and 5, including the MoTe$_2$ flake placed on top of the graphene channel, is used for spin absorption. Au/Ti contacts 4 and 7 are used to measure spin-to-charge current conversion in the MoTe$_2$ flake.

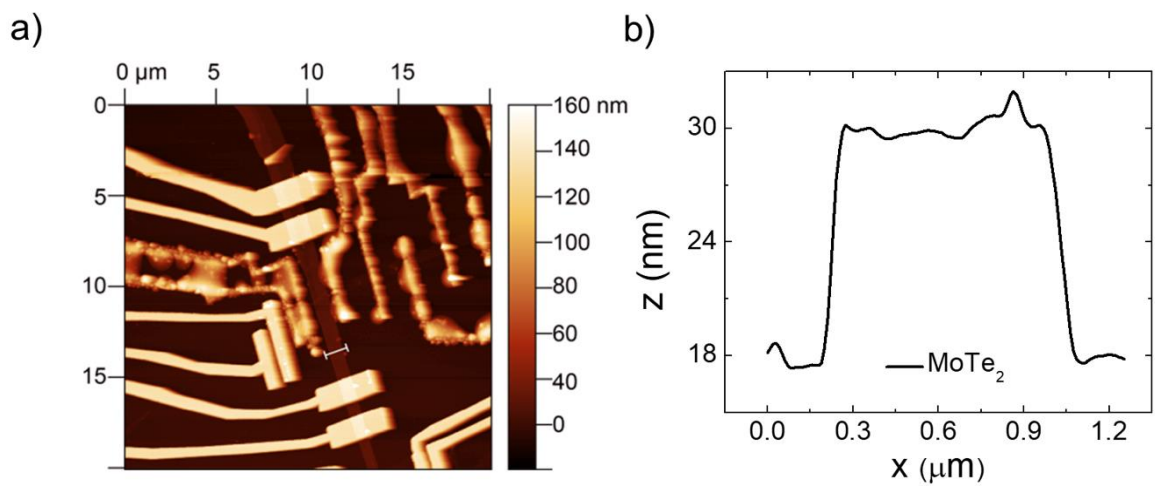

**Figure S2. Atomic force microscopy of device 1.** (a) Area scan obtained from the atomic force microcopy (AFM) technique showing the topography of device 1. The image was taken after the electrical measurement. The blurry shape of Co electrodes is caused by oxidation occurred between the measurement and the AFM characterization. (b) Line profile across the MoTe$_2$ flake along the marked line in Figure S2a. The thickness of the MoTe$_2$ flake is ~11±2 nm.

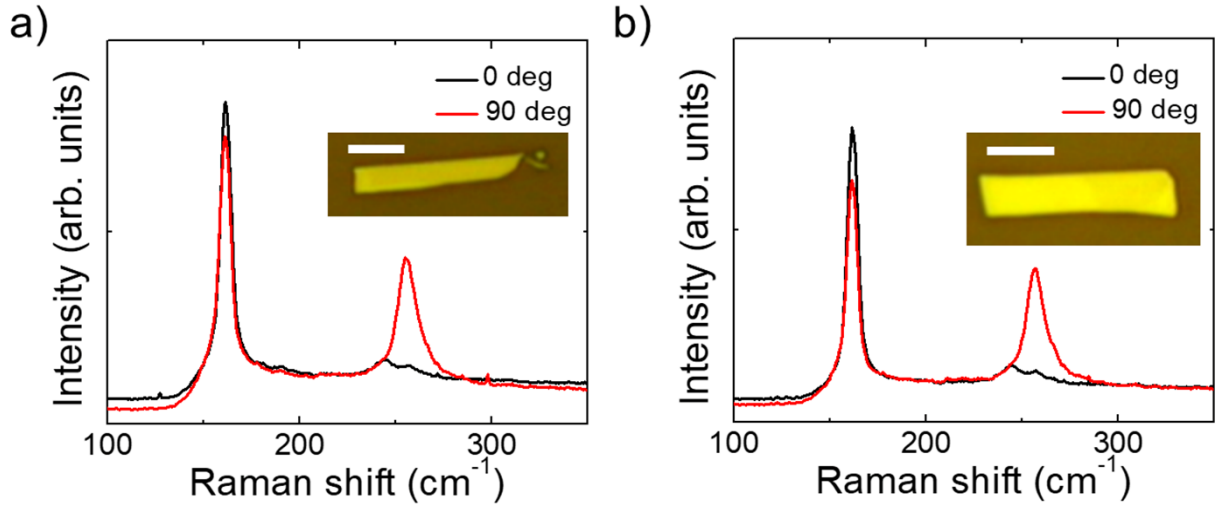

**Figure S3. Polarized Raman spectroscopy of the MoTe$_2$ needle-shaped flakes to determine the crystallographic orientation.** We performed Polarized Raman spectroscopy measurements on two different MoTe$_2$ flakes (panel a and b) with an elongated needle-like shape similar to the flake used in our device. Measurements were done with a confocal microscope using a green polarized laser (532 nm) parallel to the polarizer before the spectrometer. The 0° angle corresponds to the polarization along the long axis of the flake. Due to the maximum at ~260 cm$^{-1}$ when the angle is 90°, we can determine the crystallographic orientation of the flake[1,2]. Insets: Optical microscope image of the flakes. The scale bar of the insets corresponds to 5 µm. From these measurements, we identify that the exfoliated MoTe$_2$ flakes cleave through a preferred crystallographic axis resulting in needle-shaped flakes[3,4].

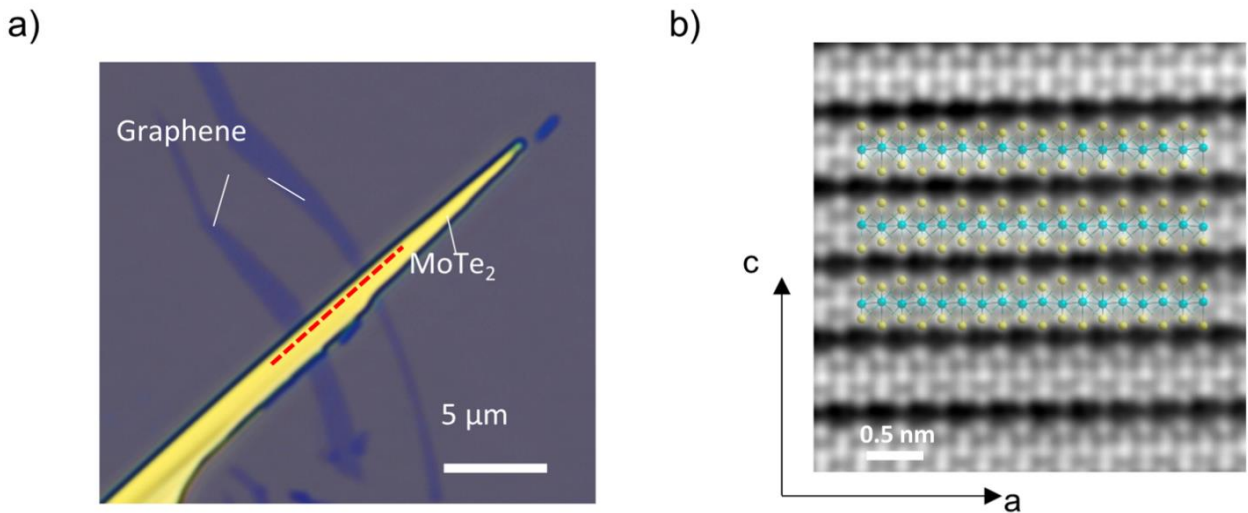

**Figure S4. Scanning transmission electron microscopy (STEM) characterization of a MoTe$_2$ needle-shaped flake to determine the crystallographic orientation.** (a) Optical image of the needle shaped MoTe$_2$ flake on top of two graphene flakes. (b) The STEM image of the flake along the red dashed line shown in Figure S4a. It shows the cleaving direction of the flake along the crystallographic axis where the zig-zag Mo-chain lies ($a$-axis). The crystallographic structure of the 1T'-MoTe$_2$, where the Mo atoms are in blue and Te atoms in dark yellow, is superimposed. The crystal orientation of the flake is in agreement with that obtained from the Raman spectroscopy measurements in Figure S3.

## S2. Temperature dependence of sheet resistance of graphene, resistivity of MoTe$_2$ and graphene MoTe$_2$ interface resistance

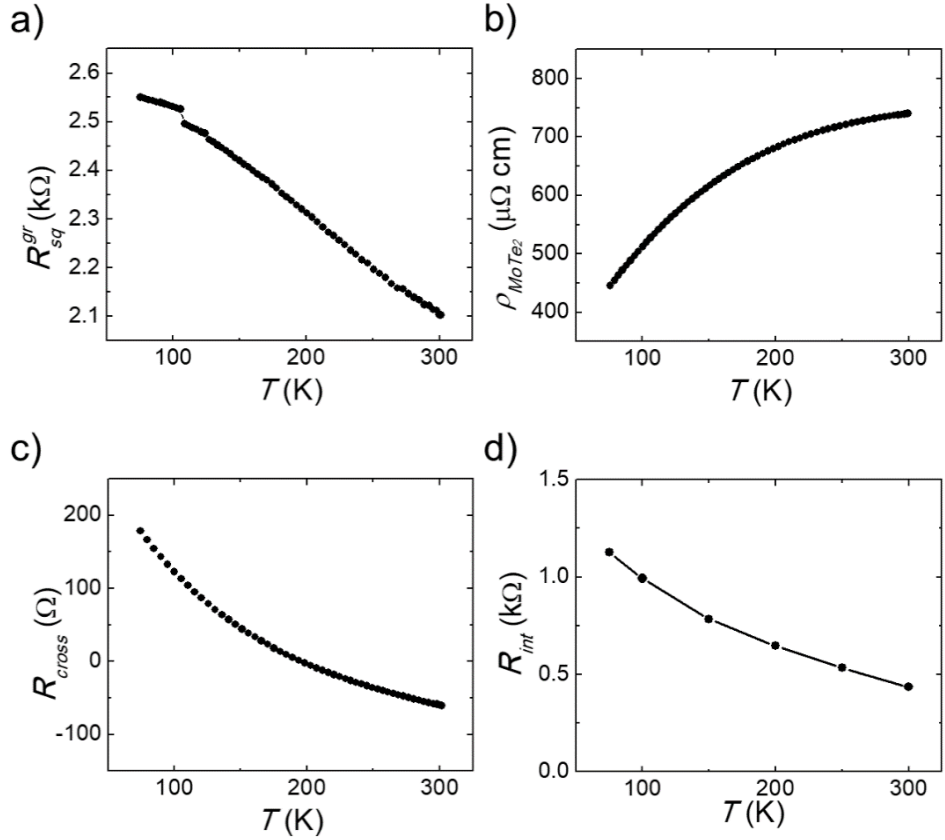

**Figure S5.** (a) Sheet resistance $R_{sq}^{gr}$ of the graphene flake used in the LSV device obtained from 4-point electrical measurements (using the terminal configuration $V_{2,3}I_{1,6}$ as shown in Figure S1) as a function of temperature. (b) Resistivity of MoTe$_2$ $\rho_{\text{MoTe}_2}$ obtained from 2-point electrical measurement ($V_{7,4}I_{7,4}$ terminal configuration as shown in Figure S1) as a function of temperature. If the material undergoes a phase transition from the 1T' to the T$_d$ phase, a jump of the resistivity with an hysteretic behavior around 240 K is expected[5]. In our case, this is not observed, evidencing that our MoTe$_2$ flake does not change phase in the range of temperatures we measure. The same effect has been observed by Stiehl et al.[3]. (c) Resistance $R_{cross}$ obtained from 4-point electrical measurements measured across the graphene/MoTe$_2$ interface (using the terminal configuration $V_{4,1}I_{7,6}$ as shown in Figure S1). (d) Interface resistance extracted using data in panels a–c. See Figure S6 for details.

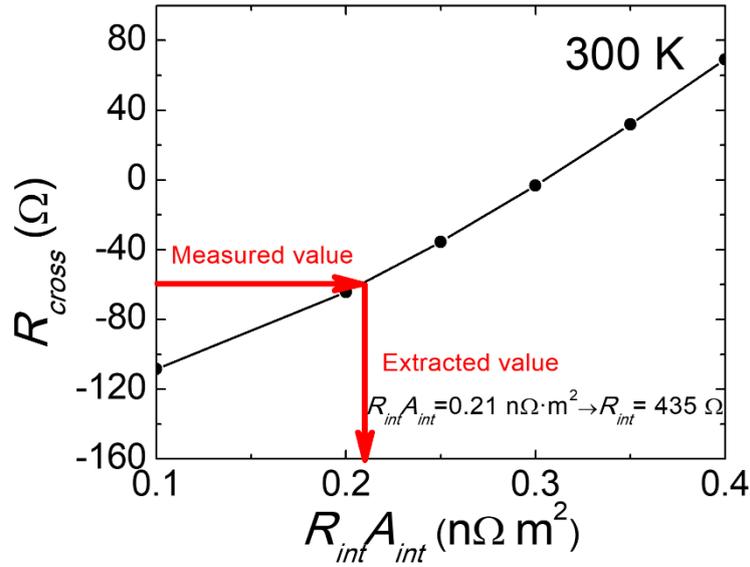

**Figure S6.** Simulation of the charge current distribution between the graphene channel and MoTe$_2$ flake using a finite element method (FEM)[6]. The geometry construction and 3D-mesh were elaborated using the free software GMSH with the associated solver GETDP[7] for calculations, post-processing and data flow control. By introducing the experimental $R_{sq}^{gr}$ and $\rho_{MoTe_2}$ values in the FEM simulation and varying the interface resistance area product ($R_{int}A_{int}$), we can recover the measured value across the interface ($R_{cross}$) for a given value of $R_{int}A_{int}$. This is done for each temperature (only the example at 300 K is shown in this figure). When $R_{int}$ is small enough, the aspect ratio between the thickness and the width plays an important role, even leading to negative values of the measured $R_{cross}$. From the FEM simulation, we also extract the shunting factor ($x_{shunt}$) needed to properly quantify the spin-to-charge conversion signal (see Note S12).

## S3. Spin transport in the reference graphene channel and easy-axis determination of the ferromagnet acting as an injector

The net $\boldsymbol{j_s}$ reaching the graphene/MoTe$_2$ interface depends on different spin transport parameters: the spin polarization of the Co/graphene interface injector and detector ($P_{inj}$ and $P_{det}$, respectively), the spin lifetime ($\tau_s^{gr}$) and the spin diffusion constant ($D_s^{gr}$) of graphene. To quantify them, we performed standard Hanle precession measurements using the reference LSV (using $V_{2,1}I_{3,6}$ terminal configuration as shown in Figure S1) as follows: First, by applying $B_y$, the initial state of the magnetizations of the two Co electrodes are fixed either parallel or antiparallel to each other. Second, for each case, $R_{NL}$ was measured by applying field along the out-of-plane hard axis ($B_z$). As explained in the main text, $B_z$ causes precession and decoherence of the spins, resulting in the oscillation and decay of the signal. In addition, the rotation of the Co magnetizations with $B_z$ tends to align $\boldsymbol{s}$ with the applied field, restoring the $R_{NL}$ signal to its zero-field value when the Co electrodes reach parallel magnetizations along the $z$-direction at saturation magnetic fields. By the combination of Hanle measurements for initial parallel ($R_{NL}^P$) and antiparallel ($R_{NL}^{AP}$) states of the Co electrodes (Figure S7), the contribution from spin precession and decoherence $\Delta R_{NL} = (R_{NL}^P - R_{NL}^{AP})/2$ can be obtained (Figure S8a). The Hanle curve for the spin precession and decoherence is then fitted to the solution of the Bloch equations using the model developed by Popinciuc et al.[8] and Maassen et al.[9]. The fitting of Hanle measurements at 300 K is shown in Figure S8a. From this fitting, the $\tau_s^{gr}$, $D_s^{gr}$ and $P = \sqrt{P_{inj}P_{det}}$ values at each temperature are obtained (Figures S8b,c and d, respectively). Note that $P_{inj}$ and $P_{det}$ cannot be obtained separately.

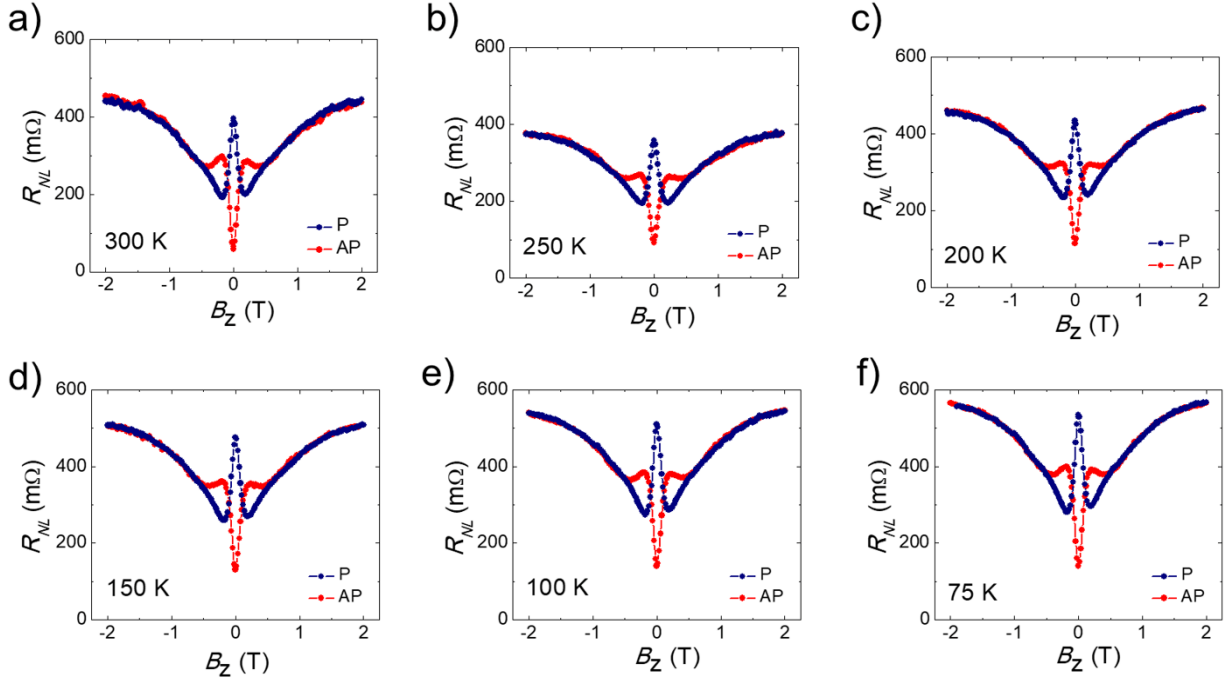

**Figure S7.** The experimental measurement of Hanle spin precession at reference LSV (using $V_{2,1}I_{3,6}$ terminal configuration shown in Figure S1) by applying $B_z$ for initial parallel ($R_{NL}^P$, blue curve) and antiparallel ($R_{NL}^{AP}$, red curve) states of the Co electrodes, at temperature (a) 300 K, (b) 250 K, (c) 200 K, (d) 150 K, (e) 100 K and (f) 75 K.

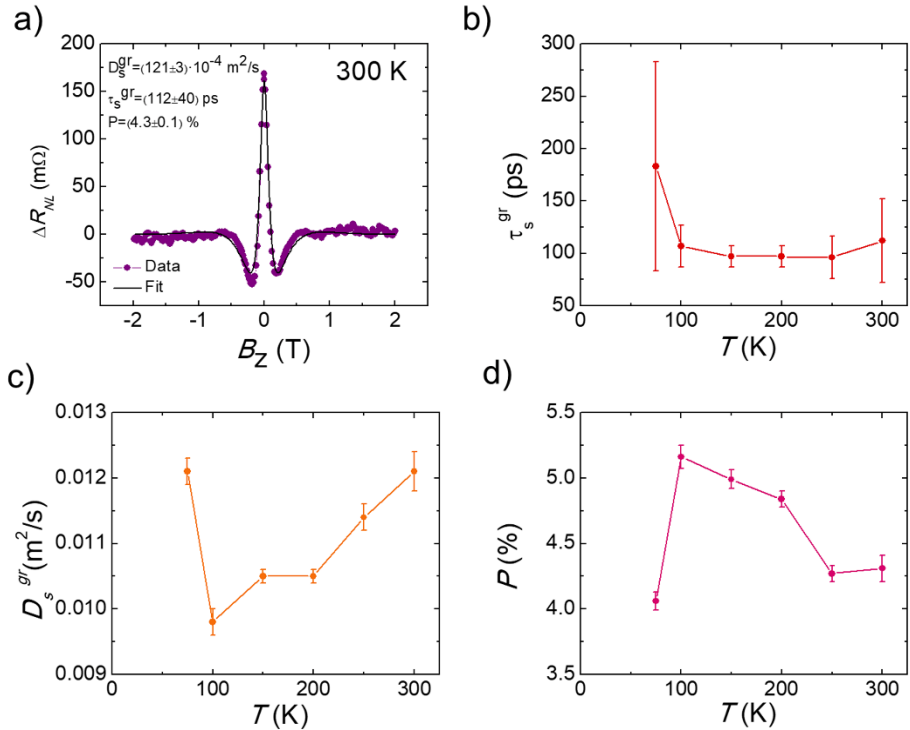

**Figure S8.** (a) The $\Delta R_{NL} = (R_{NL}^P - R_{NL}^{AP})/2$ vs. $B_z$ obtained from the experimental data in Figure S7a (purple solid circle) and the fitting of the data (black solid line) using the solutions of the Bloch equation. The plots of (b) the spin lifetime of pristine graphene ($\tau_s^{gr}$), (c) the spin diffusion constant of pristine graphene ($D_s^{gr}$), and (d) the spin polarization of the Co/graphene interface ($P = \sqrt{P_{inj}P_{det}}$) as a function of temperature.

Even though the magnetization of the ferromagnetic electrodes (with widths from 250 nm to 400 nm) should be dominated by the shape anisotropy, they could eventually present domains along the hard-axis due to the magnetocrystalline anisotropy of Co and show hysteresis loops. Since $B_x$ is applied perpendicular to the long direction of the ferromagnetic electrodes ($y$-axis), the magnetization of Co is pulled towards the direction of the field an angle $\theta_M$. Based on the conventional Hanle precession data in the parallel and antiparallel configuration shown in Figure S9a, we can determine $\theta_M$ for each value of the magnetic field. One of the terms of the Hanle precession data includes the signal due to non precessing spins, parallel to the magnetic field, which has the same sign for $R_{NL}^P$ and $R_{NL}^{AP}$ and it is proportional to $\sin^2(\theta_M)$ (see Note S12). To obtain the data plotted in Figure S9b, we have normalized $0 < R_{NL}^P + R_{NL}^{AP} < 1$ and taken the arcsine of its square root to obtain $\theta_M$ as a function of the magnetic field. Based on the values of $\theta_M$ at zero and low values of the magnetic field, we conclude that the easy axis of the magnetization is along the $y$-axis.

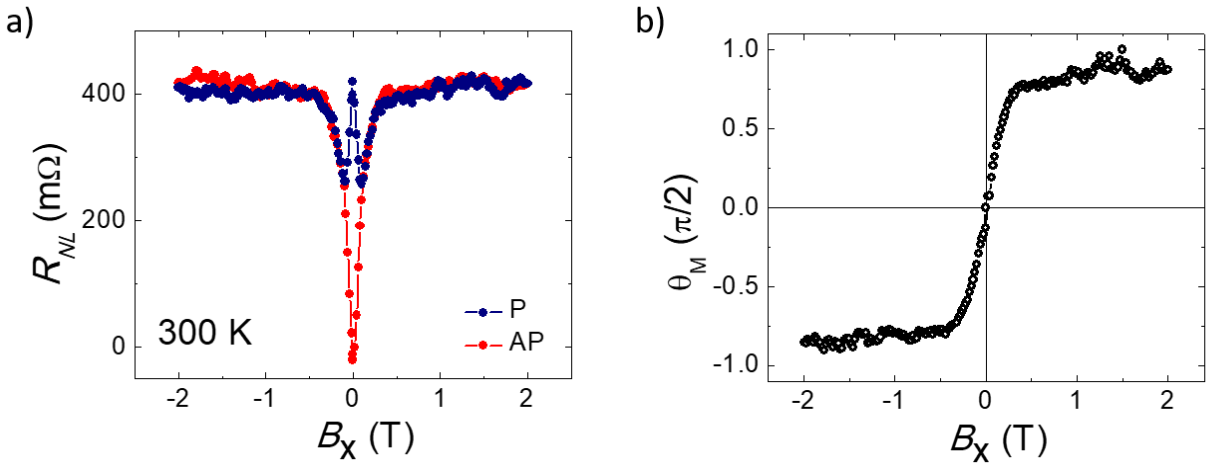

**Figure S9.** (a) Conventional Hanle precession data in the parallel (blue curve) and anti-parallel (red curve) states of the Co electrodes at 300 K. (b) Angle $\theta_M$ between the Co magnetization and the easy axis extracted from the Hanle data in panel a.

## S4. Spin absorption in MoTe₂

To verify whether the spin current is absorbed into MoTe$_2$, we compare the spin signal measured across LSVs with and without MoTe$_2$. We clearly see a decrease in the spin signal in LSV with MoTe$_2$ (Figure S10). This indicates that the spin current is strongly absorbed to MoTe$_2$ before reaching the Co detector. The net $\boldsymbol{j_s}$ reaching the MoTe$_2$ flake depends on the spin resistance of graphene ($R_s^{gr}$), the graphene/MoTe$_2$ interface resistance ($R_{int}$) and the spin resistance of MoTe$_2$ ($R_s^{MoTe_2}$). By comparing spin signals across LSVs with and without the SOC material in between, the spin diffusion length of the SOC material can be quantified[10]. However, in our case, the graphene/MoTe$_2$ interface resistance dominates the spin absorption (see Note S13 for details). Therefore, within the resolution of our measurement and calculation, it is not possible to extract the value of $\lambda_{MoTe_2}$.

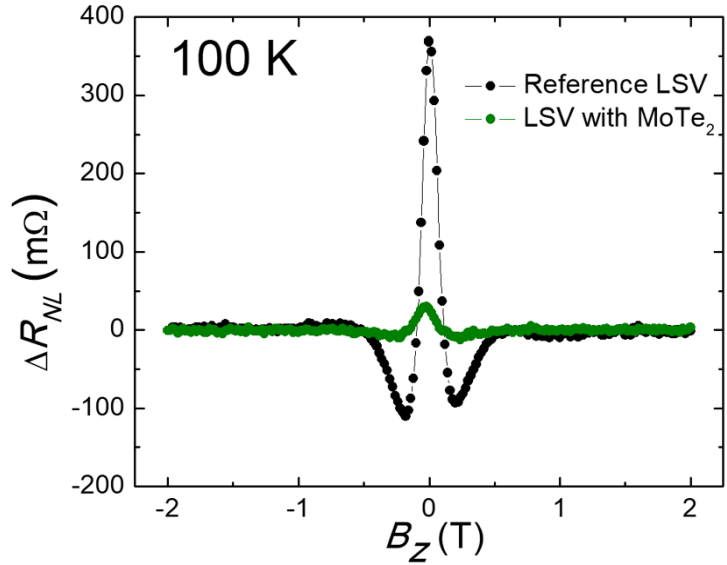

**Figure S10.** The comparison of Hanle spin precession measurement $\Delta R_{NL}$ at 100 K for the reference LSV (using $V_{2,1}I_{3,6}$ terminal configuration in Figure S1) and the LSV with MoTe$_2$ (using $V_{5,6}I_{3,1}$ terminal configuration in Figure S1) shown in black and green respectively. The spin signal for the latter case is much smaller than that the former, indicating the absorption of spin current to MoTe$_2$.

## S5. Temperature dependence of $R_{NL}$ vs. $B_x$ measurements

In the $R_{NL}$ vs. $B_x$ measurement, the symmetric component of the SCC signal (i.e., caused by $s_x$) shows an S-shaped behavior that does not depend on the initial magnetization of the Co electrode saturated along positive or negative $y$-direction (Figure 2b and 2f in the main text). However, the antisymmetric component of the SCC signals (i.e. caused by $s_y$) for the two opposite initial easy axis magnetization states of the Co, are opposite to each other (Figure 2c and 2g in the main text). Therefore, averaging the $R_{NL}^{\uparrow}$ and $R_{NL}^{\downarrow}$ data ($\frac{R_{NL}^{\uparrow}+R_{NL}^{\downarrow}}{2}$) in Figure S11 removes the signal due to the SCC with $s_y$ resulting in the S-shaped curve (antisymmetric with respect to $B_x$). The variation of this antisymmetric component of $R_{NL}$ vs. $B_x$ at different temperatures (75 K to 300 K) is shown in Figure S12a. Also, the difference between the $R_{NL}^{\uparrow}$ and $R_{NL}^{\downarrow}$ data ($\frac{R_{NL}^{\uparrow}-R_{NL}^{\downarrow}}{2}$) in Figure S11 corresponds to the contribution of SCC with $s_x$ only, resulting in the Lorentzian-shaped curve (symmetric with respect to $B_x$). The variation of this symmetric component with respect to the temperature is shown in Figure S12b. The amplitudes of both components increase with increasing temperature (see Note S8 for a detailed comparison with other measurements).

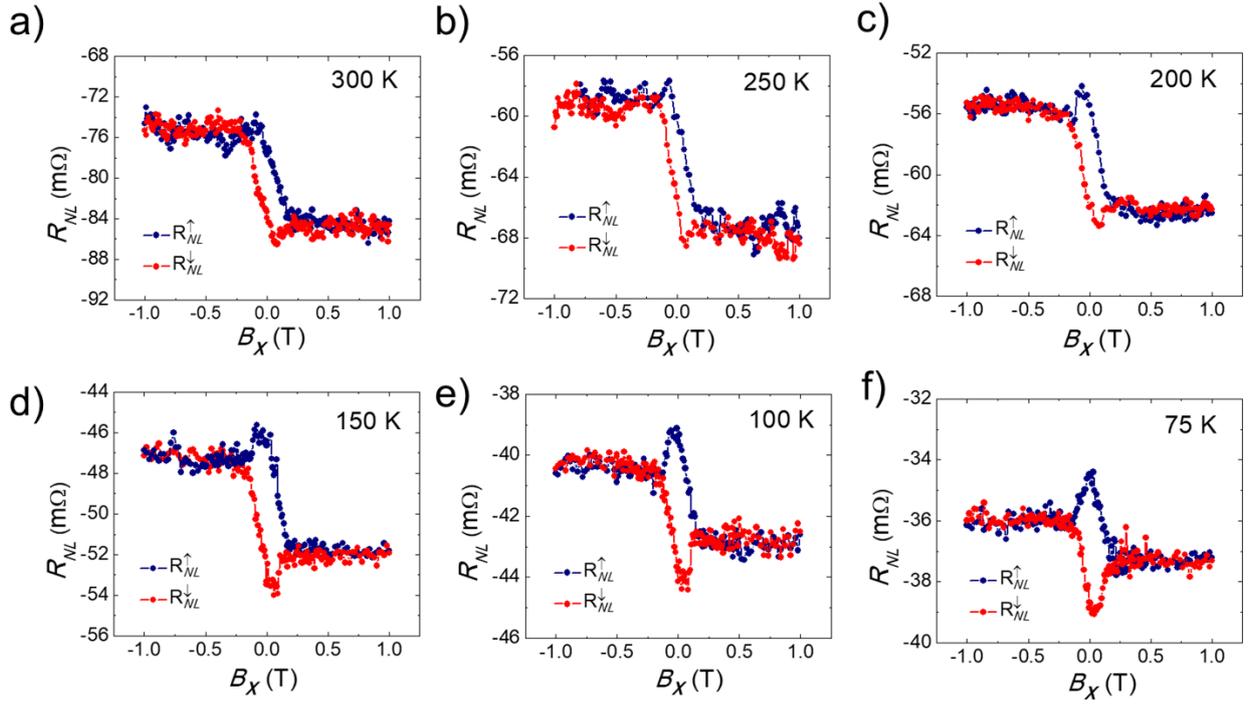

**Figure S11.** Nonlocal spin-to-charge conversion resistance ($R_{NL}$) measured (using $V_{7,4}I_{3,1}$ terminal configuration shown in Figure S1) as a function of the magnetic field applied along $x$-direction ($B_x$), i.e., the in-plane hard axis of the Co electrode, for initial magnetization of the Co electrode saturated along positive (blue) and negative (red) $y$-direction at temperature (a) 300 K, (b) 250 K, (c) 200 K, (d) 150 K, (e) 100 K, and (f) 75 K.

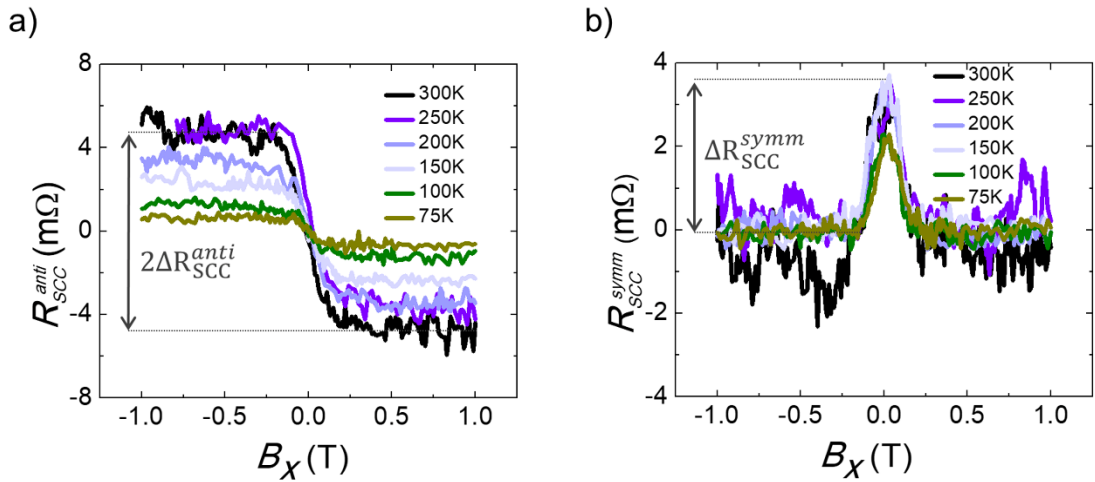

**Figure S12.** Symmetric and the antisymmetric component of the $R_{NL}$ vs. $B_x$ at different temperatures (75 K to 300 K) extracted from the data in Figure S11. (a) The average of the antisymmetric component $R_{SCC}^{anti}$. The amplitude of the signal between the two saturation states corresponds with the double of $R_{SCC}^{anti}$ amplitude (b) The average of the symmetric component $R_{SCC}^{symm}$. The amplitude of the signal from the maximum to the background corresponds to $R_{SCC}^{symm}$ amplitude signal.

## S6. Temperature dependence of $R_{NL}$ vs. $B_y$ measurements

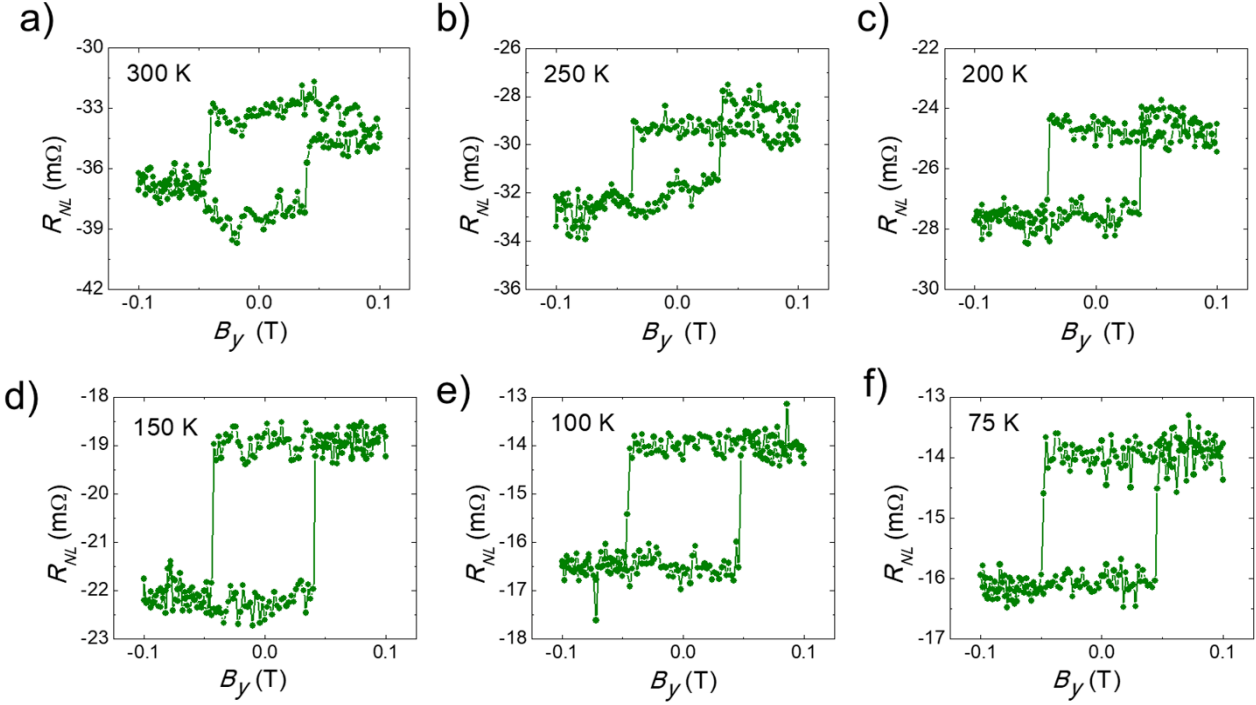

**Figure S13.** Nonlocal spin-to-charge conversion resistance ($R_{NL}$) measured (using $V_{7,4}I_{3,1}$ terminal configuration, see Figure S1a) as a function of the magnetic field applied along $y$-direction ($B_y$), i.e., the easy axis of the Co electrode, at temperature (a) 300 K, (b) 250 K, (c) 200 K, (d) 150 K, (e) 100 K and (f) 75 K. The amplitude increases with increasing temperature (see Note S8 for a detailed comparison with other measurements).

## S7. Temperature dependence of $R_{NL}$ vs. $B_z$

In the $R_{NL}$ vs. $B_z$ measurement (Figures S14a–f), the $s_x$ and $s_y$ components of the SCC signal reverse if the initial magnetization of the Co electrode switched from positive to negative $y$-direction. Therefore the average of $R_{NL}^{\uparrow}$ and $R_{NL}^{\downarrow}$ data ($\frac{R_{NL}^{\uparrow}+R_{NL}^{\downarrow}}{2}$), which is defined as $R_{SCC}$, keeps both contributions but allows to remove any baseline not related to spin signals (Figures S14g–l). The $R_{SCC}$ can be deconvoluted into the symmetric and antisymmetric part as follows:

$$R_{SCC}(B_z) = R_{SCC}^{anti}(B_z) + R_{SCC}^{symm}(B_z) = \frac{R_{SCC}(B_z) - R_{SCC}(-B_z)}{2} + \frac{R_{SCC}(B_z) + R_{SCC}(-B_z)}{2}$$

where $R_{SCC}^{anti}(B_z)$ and $R_{SCC}^{symm}(B_z)$ corresponds to SCC signal with $s_x$ and $s_y$, respectively. The curves at different temperatures are shown in Figures S14m–r, and Figures S14s–x, respectively.

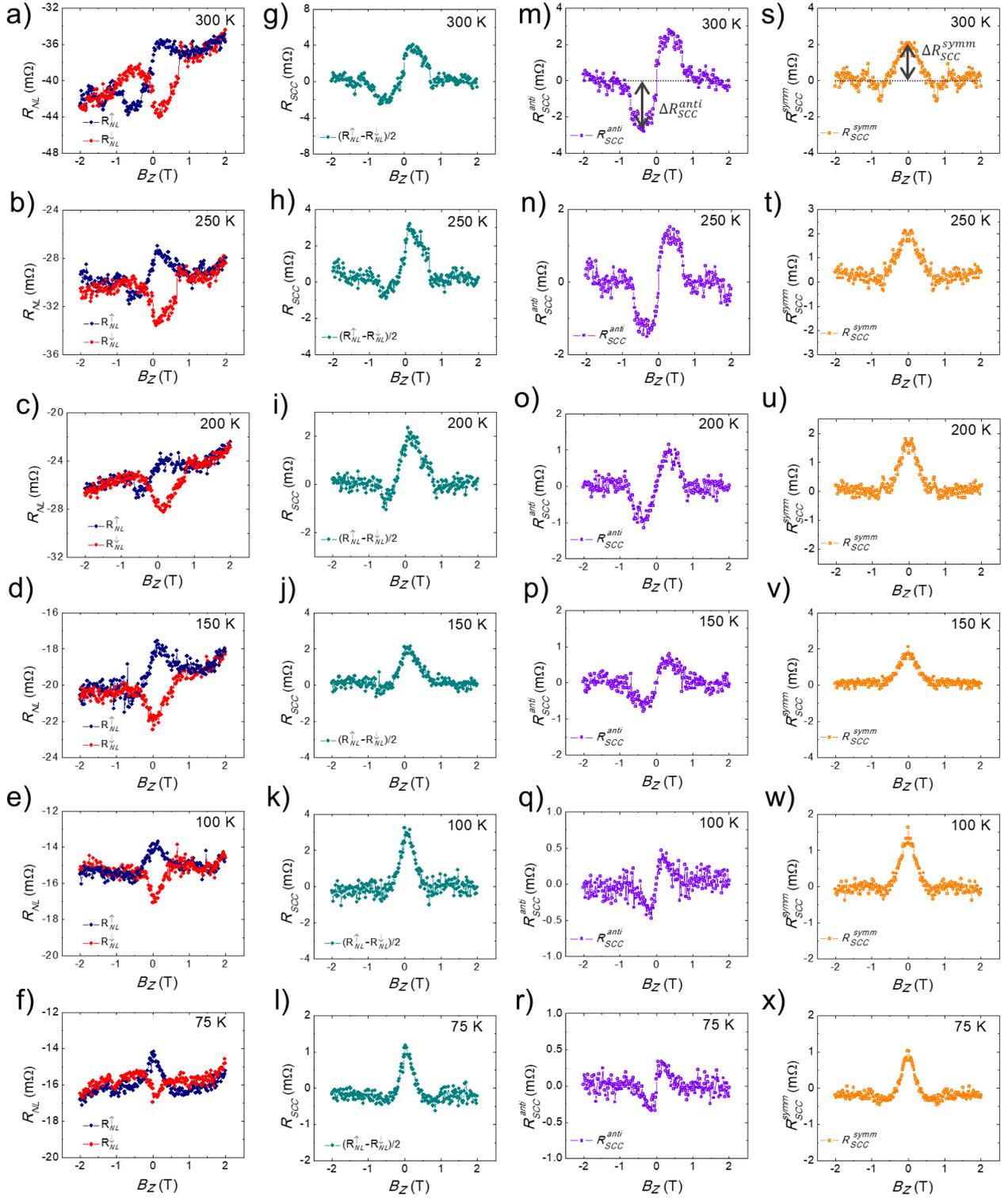

**Figure S14.** (a-f) Nonlocal spin-to-charge conversion resistance ($R_{NL}$) measured (using $V_{7,4}I_{3,1}$ terminal configuration shown in Figure S1) as a function of the magnetic field applied along $z$-direction ($B_z$), i.e., the out-of-plane hard axis of the Co electrode, for initial positive (blue) and negative (red) magnetization directions of the Co electrode at different temperatures from 300 K to 75 K. (g-l) The SCC signal $R_{SCC} = (R_{NL}^{\uparrow} - R_{NL}^{\downarrow})/2$ obtained from the data in panels a–f. (m-r) The antisymmetric component of the $R_{SCC}$ curve shown in panels g-l. It corresponds to the contribution of $s_x$ to SCC. The amplitude $\Delta R_{SCC}^{anti}$ is defined from the background to the minimum of the peak. Applying negative $B_z$ sets the spin polarization along $+x$-direction and the signal $\Delta R_{SCC}^{anti}$ is negative. The symmetric component of the $R_{SCC}$ curve shown in panels g-l. It corresponds to the contribution of $s_y$ to SCC. The amplitude $\Delta R_{SCC}^{symm}$ is defined from the background to the maximum of the peak.

## S8. Comparison of amplitudes of the SCC signal with $s_y$ and $s_x$ obtained from $R_{NL}$ vs. $B_x$, $R_{NL}$ vs. $B_y$ and $R_{NL}$ vs. $B_z$ measurements

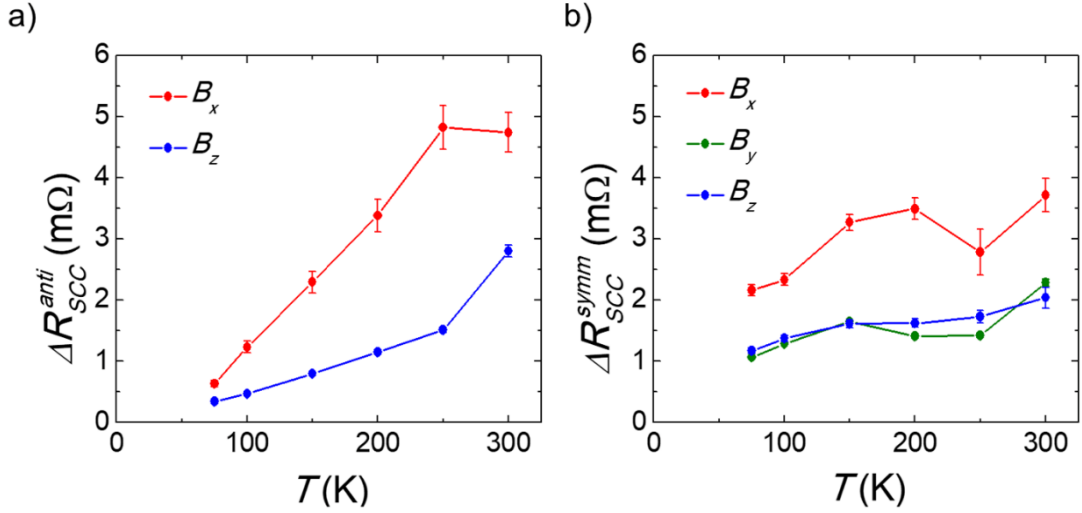

**Figure S15.** (a) Comparison of the amplitude of the antisymmetric component of the SCC (corresponding to the SCC with $s_x$) calculated from $R_{NL}$ vs. $B_x$ (red) and $R_{NL}$ vs. $B_z$ (blue) measurements. (b) Comparison of the amplitude of the symmetric component of the SCC (corresponds to the SCC with $s_y$) calculated from $R_{NL}$ vs. $B_x$ (red), $R_{NL}$ vs. $B_y$ (green) and $R_{NL}$ vs. $B_z$ (blue) measurements. For both cases, the amplitude of the signal obtained from $R_{NL}$ vs. $B_x$ measurement is larger than the other two cases. This is because the device parameters changed from the former case as the sample was taken out from the vacuum condition of the measurement system causing slight oxidation of the Co electrode and a change in the resistance of the Co/Graphene interface, leading to more efficient spin injection. However, $R_{NL}$ vs. $B_x$ measurements are used in the main text only for qualitative comparison. For the quantitative analysis (extraction of spin transport parameters and SCC efficiencies), the data obtained from the measurements with $B_z$ are used. Therefore, the change in the sample condition does not affect our analysis or interpretation.

## S9. Reproducibility

Although in the main text we focus on the results obtained in one device (device 1), we observed qualitatively similar SCC signals in another device (device 2, characterized in Figure S16), showing the robustness of the unconventional SCC signals in our graphene MoTe$_2$ van der Waals heterostructures (Figures S17-S19). A complete set of data could not be measured in this device, preventing us from quantifying the SCC efficiency in this case.

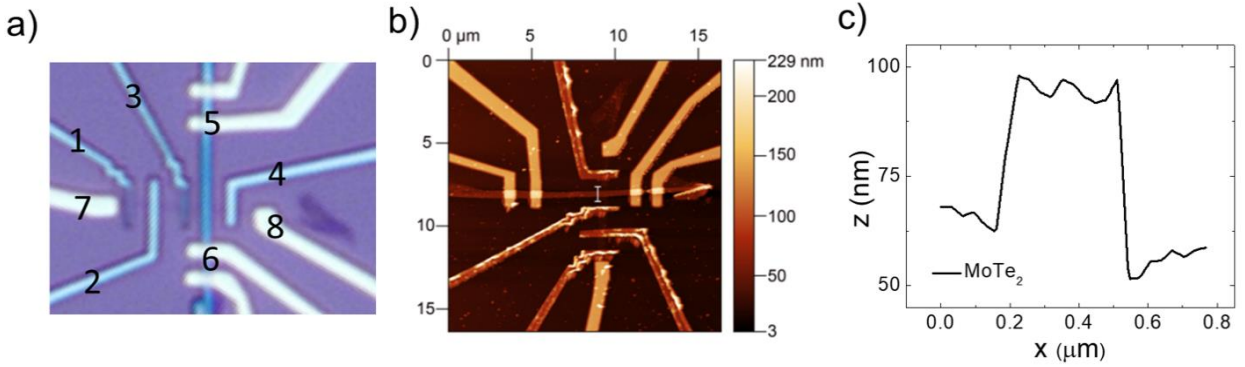

**Figure S16.** (a) Optical image of device 2. (b) Area scan showing the topography of device 2 after the electrical measurement. (c) Line profile across the MoTe$_2$ flake along the marked line in panel b. The thickness of the MoTe$_2$ flake is 40±5 nm.

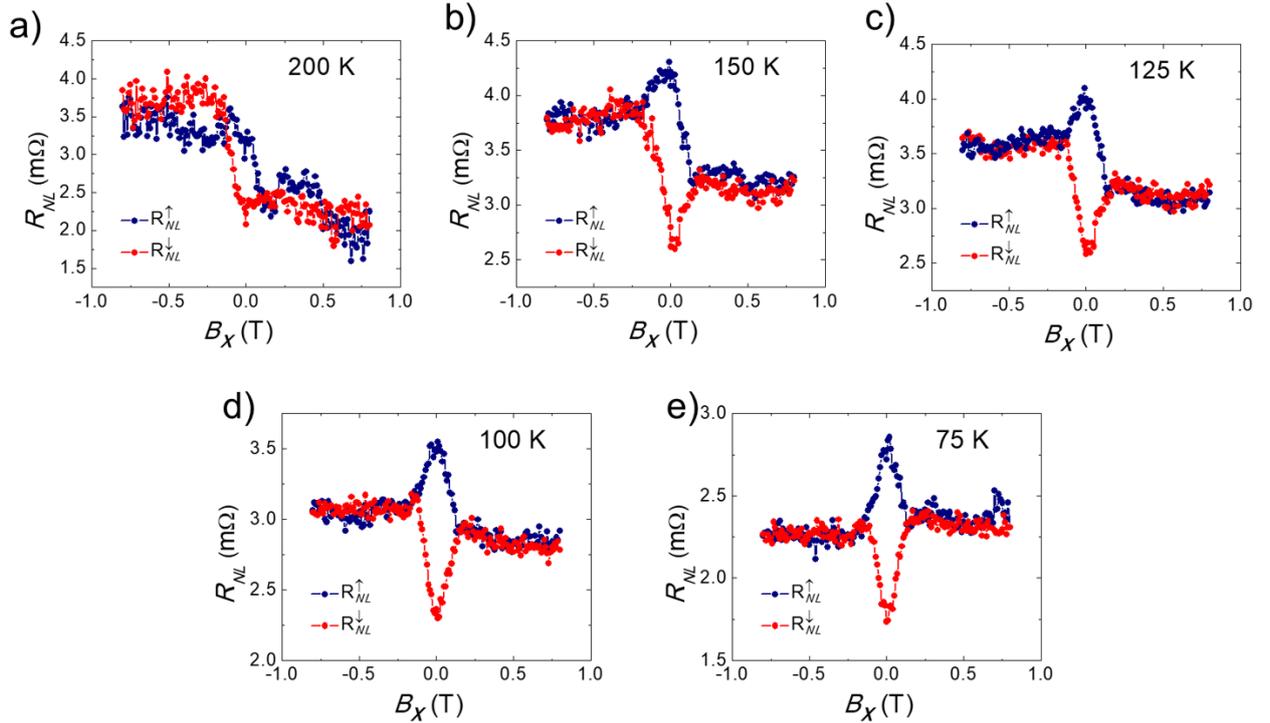

**Figure S17.** Nonlocal spin-to-charge conversion resistance ($R_{NL}$) measured (using $V_{5,6}I_{4,8}$ terminal configuration shown in Figure S16a) as a function of the magnetic field applied along $x$-direction ($B_x$), i.e., the in-plane hard axis of the Co electrode, for initial magnetization of the Co electrode saturated along positive (blue) and negative (red) $y$-direction at temperature (a) 200 K, (b) 150 K, (c) 125 K, (d) 100 K, (e) 75 K. Above 200 K, the signal to noise ratio decrease so that no spin signal could be distinguished.

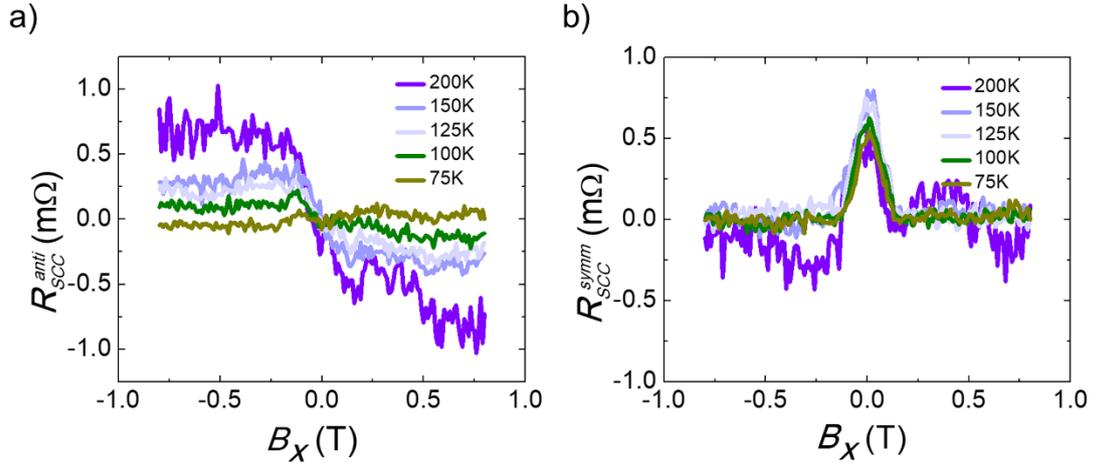

**Figure S18.** (a) Symmetric and (b) antisymmetric component of $R_{NL}$ vs. $B_x$ at different temperatures (75 K to 200 K) extracted from the data in Figure S17. These results are qualitatively similar to the measurement in device 1 (see Figure S12).

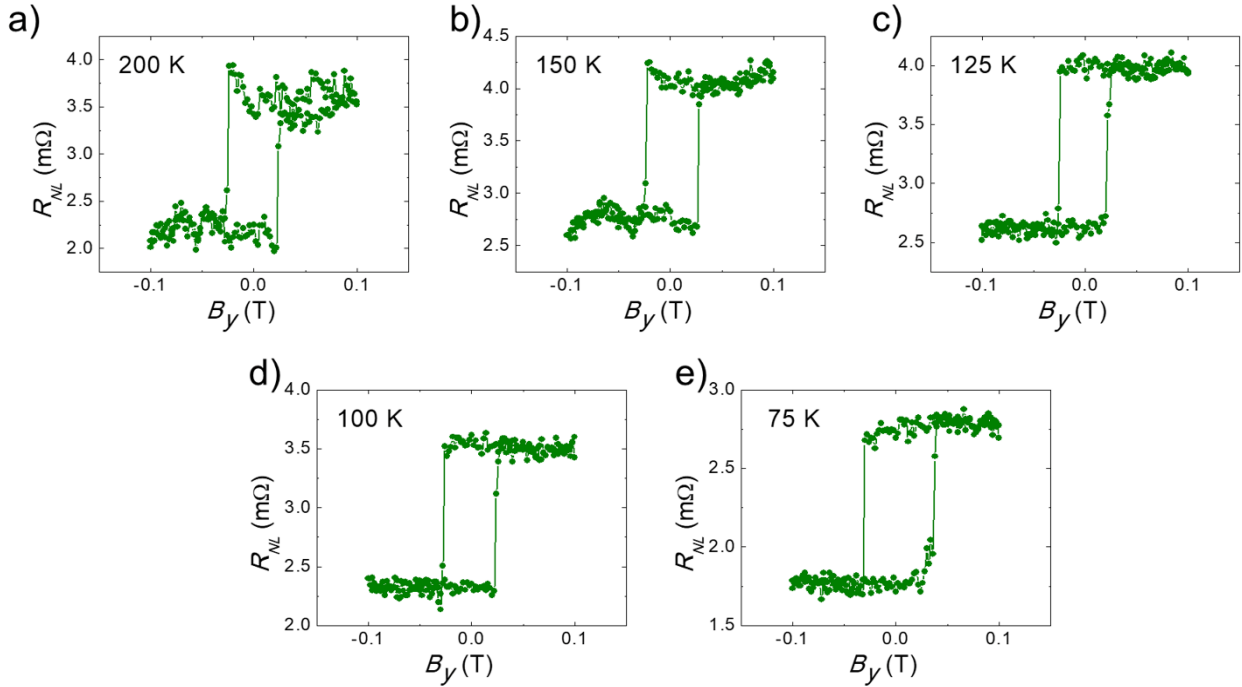

**Figure S19.** Nonlocal spin-to-charge conversion resistance ($R_{NL}$) measured (using $V_{5,6}I_{4,8}$ terminal configuration, see Figure S16a) as a function of the magnetic field applied along $y$-direction ($B_y$), i.e., the easy axis of the Co electrode, at temperature (a) 200 K, (b) 150 K, (c) 125 K, (d) 100 K and (e) 75 K. The amplitude increases with increasing temperature. These results are qualitatively similar to the measurement in device 1 (see Figure S13).

## S10. $R_{NL}$ vs. $B_y$ measurement by injecting spin current from two Co electrodes with different coercive fields.

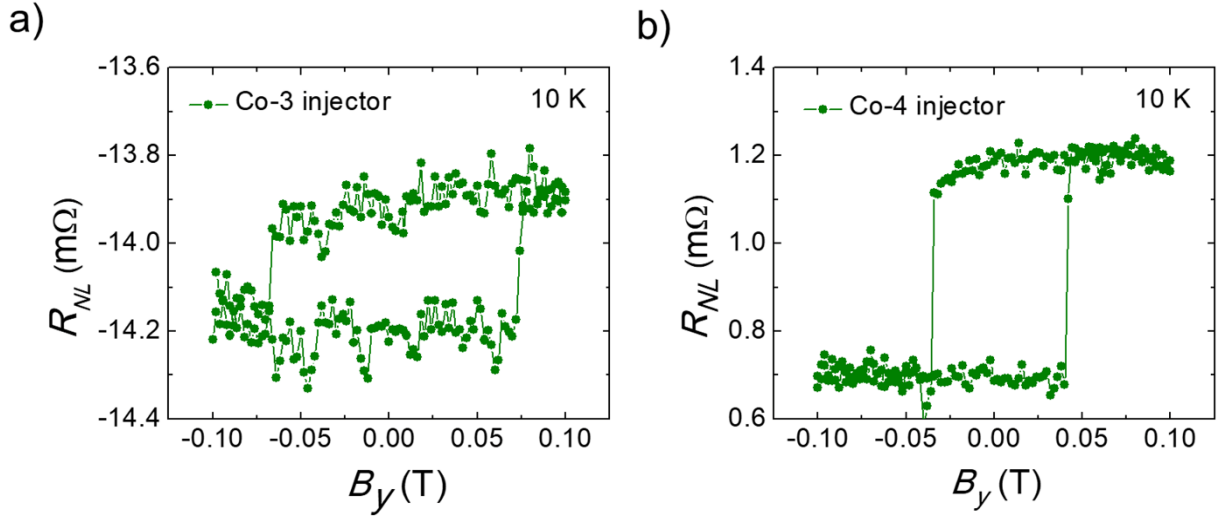

**Figure S20.** The $R_{NL}$ vs. $B_y$ measurement in device 2 at 10 K using (a) Co-3 as the injector of spin current (using $V_{5,6}I_{3,7}$ terminal configuration shown in Figure S16a) and (b) Co-4 as the injector of spin current (using $V_{5,6}I_{4,8}$ terminal configuration shown in Figure S16a). The switching of $R_{NL}$ occurs at the coercive fields of the corresponding Co injector. The width of the Co-3 electrode is narrower compared to that of Co-4. Due to shape anisotropy, the coercivity of Co-3 is larger than that of Co-4.

## S11. Onsager Reciprocity

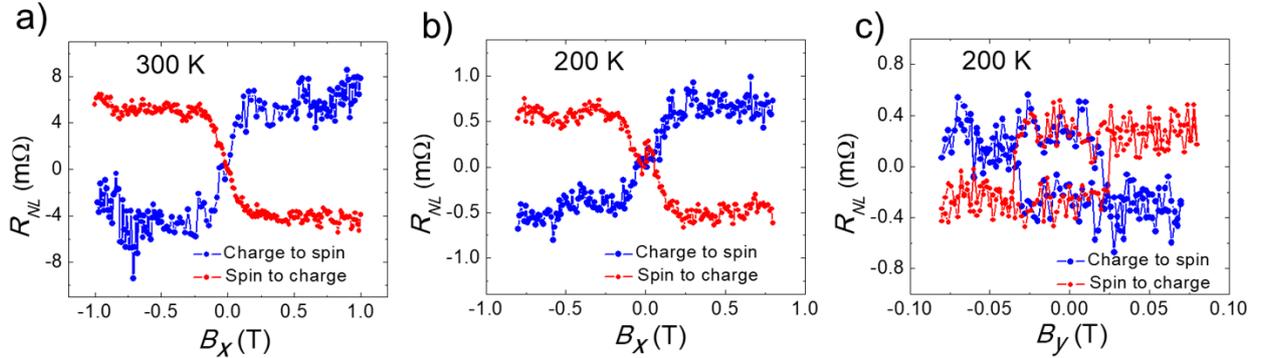

**Figure S21.** By swapping the current and the voltage terminals between Co and MoTe$_2$, we measured the spin to charge (red) and the charge to spin (blue) measurements. The antisymmetric component of the $R_{NL}$ vs. $B_x$ measurement in (a) device 1 at room temperature (here, we used electrical configuration $V_{7,4}I_{3,1}$ and $V_{3,1}I_{7,4}$ shown in Figure S16a for the spin to charge and the charge to spin measurement respectively); (b) device 2 at 200 K (here, we used electrical configuration $V_{5,6}I_{4,8}$ and $V_{4,8}I_{5,6}$ shown in Figure S16a for the spin to charge and the charge to spin measurement respectively). (c) $R_{NL}$ vs. $B_y$ measurement in device 2 at 200 K using the same electrical measurement configuration in Figure S21b. All these confirms the expected Onsager reciprocity for both $s_x$ and $s_y$ components of the SCC.

## S12. Model for spin-to-charge conversion in MoTe₂ with an interfacial barrier between the graphene and MoTe₂

To determine the spin-to-charge conversion efficiencies in MoTe₂ from the measured nonlocal signals, we model the spin propagation in our devices using the Bloch equations:

$$D_s \frac{d^2\vec{\mu}}{dx^2} - \frac{\vec{\mu}}{\tau_s} + \vec{\omega} \times \vec{\mu} = 0$$

Here $\vec{\mu} = (\mu_{s_x}, \mu_{s_y}, \mu_{s_z})$ is the spin accumulation, $D_s$ the spin diffusion coefficient and $\tau_s$ the spin lifetime. $\vec{\omega} = g\mu_B \vec{B}$ where $\vec{\omega}$ is the Larmor frequency, $g = 2$ the Landé factor, $\mu_B$ the Bohr magneton, and $\vec{B} = (B_x, B_y, B_z)$ the applied magnetic field.

When a magnetic field is applied along the $z$ direction, it induces spin precession in the $x - y$ plane. In this case, the Bloch equations have the following solution,

$$\mu_{s_y} = A\exp\left(\frac{x}{\lambda_s}\sqrt{1+i\omega\tau_s}\right) + B\exp\left(\frac{x}{\lambda_s}\sqrt{1-i\omega\tau_s}\right) + C\exp\left(-\frac{x}{\lambda_s}\sqrt{1+i\omega\tau_s}\right) + D\exp\left(-\frac{x}{\lambda_s}\sqrt{1-i\omega\tau_s}\right)$$

$$\mu_{s_x} = -iA\exp\left(\frac{x}{\lambda_s}\sqrt{1+i\omega\tau_s}\right) + iB\exp\left(\frac{x}{\lambda_s}\sqrt{1-i\omega\tau_s}\right) - iC\exp\left(-\frac{x}{\lambda_s}\sqrt{1+i\omega\tau_s}\right) + iD\exp\left(-\frac{x}{\lambda_s}\sqrt{1-i\omega\tau_s}\right)$$

where $\lambda_s = \sqrt{D_s \tau_s}$ is the spin relaxation length and $A, B, C$, and $D$ are coefficients that are determined by the boundary conditions and depend on the device geometry.

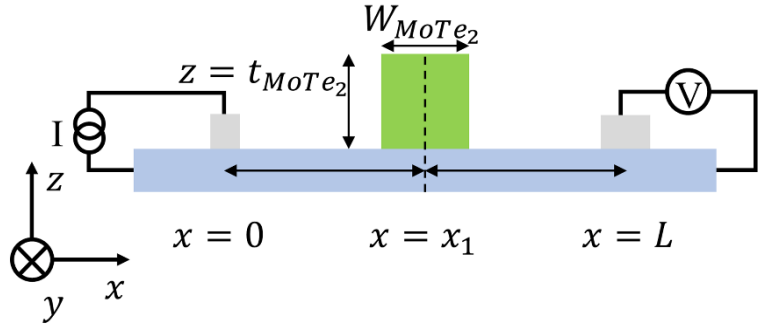

**Figure S22.** Modelled device geometry. The graphene channel is blue, and it is assumed to be infinitely long, the ferromagnetic electrodes are grey and the MoTe₂ flake is green. The non-magnetic reference electrodes are not included as they do not influence the spin transport.

In the MoTe₂ flake, because of the large spin-orbit coupling, the spin lifetime is expected to be very short. Therefore, we assume that $\omega\tau_s^{MoTe_2} \ll 1$ and we write the spin accumulation in the $x$ and $y$ directions as follows:

$$\mu_{s_y} = E\exp\left(\frac{z}{\lambda_s^{MoTe_2}}\right) + F\exp\left(-\frac{z}{\lambda_s^{MoTe_2}}\right)$$

$$\mu_{s_x} = G \exp\left(\frac{z}{\lambda_s^{MoTe_2}}\right) + H \exp\left(-\frac{z}{\lambda_s^{MoTe_2}}\right)$$

Here the coefficients $E, F, G$, and $H$ are determined by spin precession in the graphene channel and the confinement induced by the finite thickness of the MoTe$_2$ flake. Note that the lack of spin precession removes the coupling between spins in the x and y direction in this region.

As shown in Figure S22, the modeled device geometry is divided in 5 different regions:

1. The left side of the spin injector. This region is semi-infinite, extending from $x = 0$ to $x \to -\infty$ and the spin lifetime and diffusivity in this region are $\tau_s^{gr}$, $D_s^{gr}$, and the square resistance is $R_{sq}^{gr}$.
2. The region between the spin injector (placed at $x = 0$ and with contact resistance $R_{c1}$) and the MoTe$_2$ flake, that is placed at $x = x_1$ and has a width $W_{MoTe_2}$ that is assumed to be much shorter than the spin relaxation length in the graphene channel $\lambda_s^{gr} = \sqrt{\tau_s^{gr} D_s^{gr}}$. The interface between both materials has a resistance $R_{int}$.
3. The region between the MoTe$_2$ flake ($x = x_1$) and the spin detector which is placed at $x = L$ and has a contact resistance $R_{c2}$. This region has the same spin transport properties as regions 1 and 2.
4. The right side of the spin detector. This region is also semi-infinite, it extends from $x = L$ and $x \to \infty$ and has the same spin transport properties as regions 1, 2, and 3.
5. The MoTe$_2$ region extends from $z = 0$ until $z = t_{MoTe_2}$, connecting with regions 2 and 3 at $x = x_1$. Spin transport in region 5 is assumed to be perpendicular to the graphene plane, the spin relaxation length is $\lambda_s^{MoTe_2}$, the spin lifetime is assumed to be short enough so that spin precession is irrelevant, its resistivity is $\rho_{MoTe_2}$, and it has a finite thickness $t_{MoTe_2}$.

To determine the spin accumulations and currents in the device, we use the following boundary conditions:

1. The spin accumulation $\mu_s \to 0$ when $x \to \pm\infty$.
2. The spin accumulation is continuous everywhere apart from at the interface between the graphene and the MoTe$_2$, where it has a discontinuity which is equal to $\Delta \mu_s = eR_{int} I_s^{int}$ where $I_s^{int}$ is the spin current crossing the interface and $e$ the electron charge.
3. There is no spin relaxation at the graphene/MoTe$_2$ interface. This implies that $I_s^{int}$ is equal to the "bulk" spin current at the MoTe$_2$ at $z = 0$.
4. The spin currents are defined as $I_{s_{x(y)}}^{gr} = -\frac{W_{gr}}{eR_{sq}^{gr}} \frac{d\mu_{s_{x(y)}}}{dx}$ for spins pointing in the $x(y)$ direction in the graphene and $I_{s_{x(y)}}^{MoTe_2} = -\frac{W_{gr} \times W_{MoTe_2}}{e\rho_{MoTe_2}} \frac{d\mu_{s_{x(y)}}}{dz}$ in the MoTe$_2$ channel. $W_{gr}$ is the width of the graphene channel.
5. $I_s^{gr}$ has a discontinuity at $x = 0$ of $\Delta I_{s_y}^{gr} = P_{inj}I - \frac{\mu_{s_y}(x=0)}{eR_{c1}}$ for spins pointing along $y$ and $\Delta I_{s_x}^{gr} = -\frac{\mu_{s_x}(x=0)}{eR_{c1}}$ for spins along $x$. At the graphene/MoTe$_2$ interface, $\Delta I_{s_{x(y)}}^{gr} = -\frac{\mu_{s_{x(y)}}^{gr}(x=x_1) - \mu_{s_{x(y)}}^{MoTe_2}(z=0)}{eR_{int}}$. Here, $P_{inj}$ is the spin polarization of the spin injector and $I$ is the applied charge current. Finally, at the spin detector, $\Delta I_{s_{x(y)}}^{gr} = -\frac{\mu_{s_{x(y)}}(x=L)}{eR_{c2}}$.

The spin signal at $x = L$ is determined using: $R_S(x = L) = \pm \frac{P_{det}\mu_{s_y}(x=L)}{eI}$ where the $\pm$ stands for the parallel and antiparallel configurations and $P_{det}$ is the spin polarization of the detector. To fit our data, we have considered that the contact magnetizations get pulled along the **B** field direction by using the following formula:

$$R_{NL}^{P(AP)} = +(-)R_S(x = L)\cos^2\theta_M + R_\parallel \sin^2\theta_M.$$

Here, $R_\parallel$ is the signal induced by non-precessing spins parallel to **B** and $\theta_M$ is the contact magnetization angle with respect to the easy axis and is determined using

$$\sin^2\theta_M = \frac{f(B) - \min(f(B))}{\max(f(B)) - \min(f(B))}.$$

Here, $f(B) = R_{NL}^P(B) + R_{NL}^{AP}(B)$. We remove $R_\parallel$ from our analysis by defining the spin signal as $\Delta R_{NL} = (R_{NL}^P - R_{NL}^{AP})/2$ and fitting it to the analytical expression for $R_S(x = L)\cos^2\theta_M$ (Figure S23d). From this fitting, the $\tau_s^{gr}$, $D_s^{gr}$ and $P = \sqrt{P_{inj}P_{det}}$ values can be extracted. Note that $P_{inj}$ and $P_{det}$ cannot be obtained separately.

The $x$ and $y$ components of the spin-to-charge conversion signal are obtained using:

$$R_{SH}^{x(y)} = \pm\theta_{zy}^{x(y)}\rho_{\text{MoTe}_2}x_{shunt}\bar{I}_{s_{x(y)}}/W_{\text{MoTe}_2}.$$

Here the $\pm$ stands for the up and down magnetization configurations of the spin injector, $\theta_{zy}^{x(y)}$ is the spin Hall angle associated with the conversion of spins pointing along $x(y)$, $x_{shunt}$ is the shunting factor associated with the role of the graphene as a parallel channel that reduces the effective resistance of the MoTe$_2$ flake and it has been calculated in Note S2. $\bar{I}_{s_{x(y)}}$ is the average spin current in the MoTe$_2$ flake and is calculated using: $\bar{I}_{s_{x(y)}} = \frac{1}{t_{\text{MoTe}_2}}\int_0^{t_{\text{MoTe}_2}} I_{s_{x(y)}}(z)dz$.

In this case, because only one Co contact is involved in the measurement, we have considered its magnetization pulling using:

$$R_{NL}^{\uparrow(\downarrow)} = +(-)R_{SH}^{x(y)}\cos\theta_M + R_{SH}^\parallel \sin\theta_M.$$

Here, $R_{SH}^\parallel$ is the SCC signal induced by spins pointing along **B**. Note that, in our case, when $B_z$ is applied, $R_{SH}^\parallel = 0$, indicating that SCC along $z$ is smaller than our noise level. Nevertheless, to avoid background-related issues, we have analyzed the SCC signals using $R_{SCC} = (R_{NL}^\uparrow - R_{NL}^\downarrow)/2$. Finally, to determine the $y$-component of SCC, we have symmetrized $R_{SCC}$ and, to determine the $x$-component, we have antisymmetrized $R_{SCC}$ with respect to $B_z$. The resulting data is fit to $R_{SCC}^{anti(symm)} = R_{SH}^{x(y)}\cos\theta_M$ (Figures S24a,b), where we use $\theta_M$ extracted from the reference Hanle precession measurements as described above in this Note.

## S13. Role of the interface resistance between the graphene and MoTe$_2$ on the spin signal

To understand the role of $R_{int}$ on the spin signal and the spin-to-charge conversion, we have used our model with some typical spin transport parameters very close to our device (see Table S1) to determine the spin signal as a function of $R_{int}$. Here, it is useful to define the spin resistance of the MoTe$_2$ $R_s^{MoTe2} = \rho_{MoTe_2} \lambda_s^{MoTe_2}/(W_{MoTe_2} W_{gr} \tanh(t_{MoTe_2}/\lambda_s^{MoTe_2}))$ and of the graphene channel $R_s^{gr} = R_{sq}^{gr} \lambda_s^{gr}/W_{gr}$, which is the resistance spins experience before relaxing in the MoTe$_2$ and graphene channels, respectively. Note that, with these definitions, the graphene channel length is much longer than $\lambda_s^{gr}$, but $t_{MoTe_2}$ can be comparable to $\lambda_s^{MoTe_2}$.

| $D_S^{gr}$ $(m^2/s)$ | $\tau_s^{gr}$ (ps) | $R_{sq}^{gr}$ ($\Omega$) | $\rho_{MoTe_2}$ ($\Omega$m) | $\lambda_s^{MoTe_2}$ (nm) | $L$ ($\mu m$) | $W_{gr}$ ($\mu m$) | $W_{MoTe_2}$ ($\mu m$) | $t_{MoTe_2}$ (nm) | $x_1$ ($\mu m$) | $P$, $\theta_{SH}$ |
|---|---|---|---|---|---|---|---|---|---|---|
| 0.01 | 100 | 2500 | $7.4 \cdot 10^{-6}$ | 1 or 100 | 2 | 0.59 | 0.81 | 11 | 0.93 | 0.1 |

**Table S1.** Spin transport parameters used to illustrate the role of the interface resistance on the spin-to-charge conversion in the graphene-MoTe$_2$ devices. To simplify the picture, we assume that $R_{c1}$ and $R_{c2}$ are much larger than $R_s^{gr}$. As a consequence, they do not play any significant role in our model.

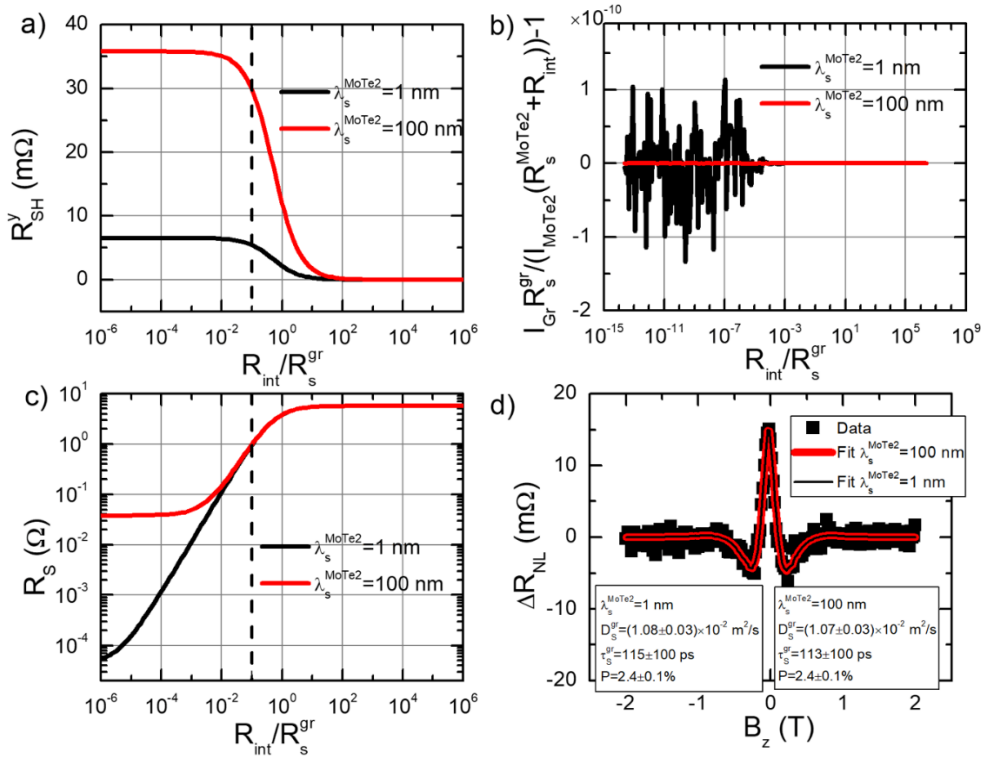

**Figure S23.** The effect of a contact resistance between the graphene and MoTe$_2$ on a) the spin-to-charge conversion signal induced by spins pointing along $y$ ($R_{SH}^y$) and c) the spin signal ($R_S$) measured at $x = L$ (Table S1). Both signals are calculated for zero magnetic field and for $\lambda_s^{MoTe_2} = 1$ and 100 nm. The gray dotted line is the interface resistance measured in device 1 at room temperature and, because $R_s^{MoTe2} < R_{int}$, $\lambda_s^{MoTe_2}$ does not play any significant role in the spin signal within the calculated range, thus preventing us to extract $\lambda_s^{MoTe_2}$. In b) we show the normalized ratio between the spin current remaining in the graphene at $x = x_1$ and the spin current absorbed in the MoTe$_2$. To show that this ratio is determined by the spin resistances and the interface resistance, we have normalized this ratio by the corresponding spin resistances. Because we observe that it is constant within a numerical error of $1 \times 10^{-10}$, we can confirm the accuracy of our calculations, also from the numerical perspective. Finally, in d) we show a fit to the Hanle precession data measured at 100 K across the

MoTe$_2$ (see Figure S7) for $\lambda_s^{MoTe_2} = 1$ and 100 nm. Because the spin absorption is dominated by $R_{int}$, there is no relevant effect of changing $\lambda_s^{MoTe_2}$ and it cannot be determined.

## S14. Quantification of the spin-to-charge conversion signal

To be able to extract the SCC efficiencies of our device, because we cannot extract $\lambda_s^{MoTe_2}$ from our experiments, we have fit the SCC data to the model described in S12 for 1 nm $< \lambda_s^{MoTe_2} <$ 20 nm. To reduce the amount of fitting parameters, we have fixed $\tau_s^{gr}$ and $P_{inj}$ (assuming $P = P_{inj} = P_{det}$) to the values extracted from the reference Hanle fitting (Figure S8) and released $D_s^{gr}$ and $\theta_{zy}^{x(y)}$. Since the sign of $P_{inj}$ is not known, the absolute sign of $\theta_{zy}^{x(y)}$ cannot be determined  In Figure S24, we show the result from such a fit at room temperature for the symmetric ($\theta_{zy}^{y}$) (a) and antisymmetric ($\theta_{zy}^{x}$) (b) cases. The $\lambda_s^{MoTe_2}$-dependence of $\theta_{zy}^{x(y)}$ c) and $\theta_{zy}^{x(y)}\lambda_s^{MoTe_2}$ d). In c) we observe how, for $\lambda_s^{MoTe_2} < t_{MoTe_2}$, $\theta_{zy}^{x(y)}$ decreases linearly with increasing $\lambda_s^{MoTe_2}$ leading to a constant $\theta_{zy}^{x(y)}\lambda_s^{MoTe_2}$. In contrast, when $\lambda_s^{MoTe_2} \sim t_{MoTe_2}$, $\theta_{zy}^{x(y)}$ becomes independent of $\lambda_s^{MoTe_2}$ and $\theta_{zy}^{x(y)}\lambda_s^{MoTe_2}$ increases linearly. This occurs because $\bar{I}_{sx(y)}$ in this range becomes almost independent of $\lambda_s^{MoTe_2}$. We have taken $\theta_{zy}^{x(y)}(\lambda_s^{MoTe_2} = 20$ nm$)$ as the lower limit for $\theta_{zy}^{x(y)}$ and $\theta_{zy}^{x(y)}\lambda_s^{MoTe_2}(\lambda_s^{MoTe_2} = 1$ nm$)$ as the lower limit for $\theta_{zy}^{x(y)}\lambda_s^{MoTe_2}$.

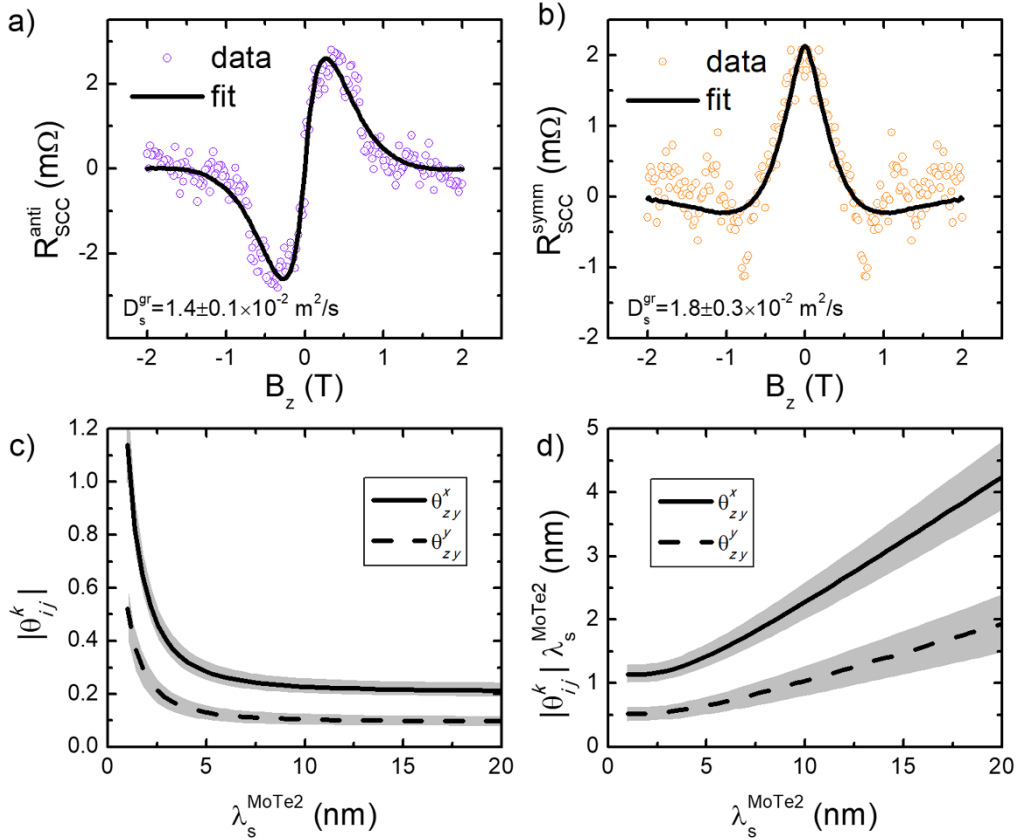

**Figure S24.** Analysis of the SCC data as a function of $\lambda_s^{MoTe_2}$ at 300 K. (a) Antisymmetrized and (b) symmetrized SCC signals corresponding to the $s_y$ and $s_x$ components of SCC, respectively, together with their fits to the model described in Note S12. (c) and (d) show $\theta_{zy}^{x(y)}$ and $\theta_{zy}^{x(y)}\lambda_s^{MoTe_2}$, respectively, as a function of $\lambda_s^{MoTe_2}$. The grey areas in (c) and (d) correspond to the uncertainties coming from the fitting errors.

## S15. In-plane angle dependence of $R_{NL}$

Since the sign of $R_{NL}$ vs $B_x$ is reversed to that of $R_{NL}$ vs $B_y$ (see Figures S12a and S13, respectively), we can obtain the value of the in-plane angle $\theta_M$ of the magnetization of the Co injector at which the voltage generated across MoTe$_2$, and hence the value of $R_{NL}$, is zero.

Considering a high enough in-plane magnetic field with an angle $\theta_M$ with respect to the easy-axis of the ferromagnet ($y$-axis), the magnetization of the ferromagnet will be aligned with the magnetic field. For this case, the non-local resistance will be given by $R_{NL} = R_{SH}^x \sin\theta_M + R_{SH}^y \cos\theta_M$, where $R_{SH}^{x(y)}$ stands for the $s_x$ and $s_y$ contributions of the spin-to-charge conversion signal. When we apply an out-of-plane magnetic field $B_z$, we can extract, in the same experiment, the values of the contribution of the symmetric and antisymmetric part corresponding to $s_y$ and $s_x$, respectively. The case of $R_{SH}^x$ will correspond to $\Delta R_{SCC}^{anti}$ (see Figure S14m), whereas the case of $R_{SH}^y$ will correspond to $\Delta R_{SCC}^{sym}$ (see Figure S14s). At room temperature, the values are $R_{SH}^x = \Delta R_{SCC}^{anti} = -2.8$ m$\Omega$ and $R_{SH}^y = \Delta R_{SCC}^{sym} = 2.0$ m$\Omega$. Note the negative sign of $R_{SH}^x$: For positive magnetic field $B_z$, the sign of $s_x$ is set by spin precession, which is proportional to the cross product of $B_z$ and the initial spin polarization along $+y$-axis ($B_z \times s_y$). Here, $-B_z$ sets the spin polarization along $+x$-direction and the output signal $\Delta R_{SCC}^{anti}$ is negative (see Figure S14m). The same result is observed for positive magnetic field $B_x$ above saturation (corresponding to spin polarization along $+x$-direction), where the signal is negative (see Figure S12a). On the other hand, $R_{SH}^y$ has a positive sign: At zero magnetic field, the sign of $s_y$ is set by the initial magnetization of the Co electrode and the output signal $\Delta R_{SCC}^{symm}$ is positive with the spin polarization along $+y$-direction (see Figure 12b and 14s). The same result is observed for positive magnetic field $B_y$ above saturation (corresponding to spin polarization along $+y$-direction), where the signal is positive (see Figure S13).

Finally, we plot the angular dependence of the non-local signal $R_{NL}$ in Figure S25, where we observe angles $\theta_{M_n}^0$ (corresponding to $\theta_{M_n}^0 = 36° + n \cdot 180°$, with $n \epsilon \mathbb{Z}$) at which the signal is zero.

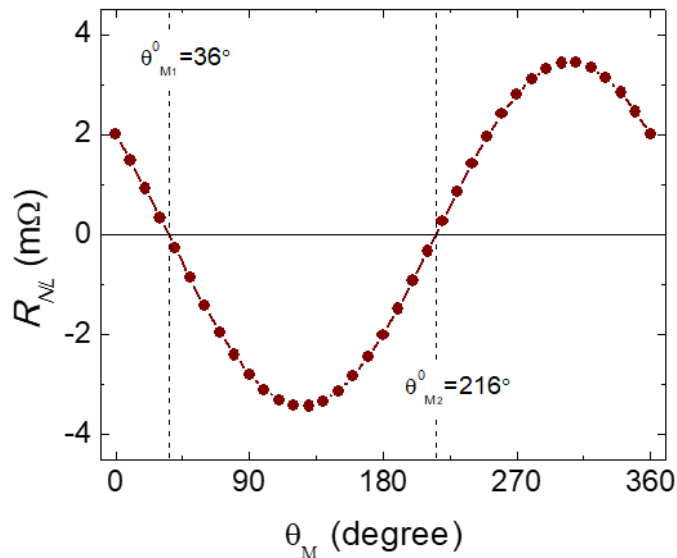

**Figure S25.** Calculated non-local resistance as a function of the angle of the in-plane magnetic field using the experimental SCC signals for the $s_x$ and $s_y$ contributions obtained at room temperature. The dashed lines correspond to the angles $\theta_{M_n}^0$ where the signal is zero.

## S16. Symmetry considerations for spin-to-charge interconversion and Density Functional Theory (DFT) calculations

As discussed in the main text, the mirror symmetry of the 1T' MoTe$_2$ structure prevents any spin accumulation where the spin points in the same direction as the applied current, regardless of whether this is produced by bulk spin Hall effect or surface Edelstein effect. It has recently been proposed that such a spin accumulation is possible in WTe$_2$ through the surface Edelstein effect due to the particular spin polarization of the Fermi arcs. However, the surfaces of 1T$_d$ WTe$_2$ and 1T' MoTe$_2$ have the same symmetry (only a remaining mirror plane), and therefore the effect is still not possible.

To explain our observations, the mirror symmetry must be broken. Since the MoTe$_2$ samples are deposited on a substrate, we speculate that this might induce shear strain that breaks the mirror. To illustrate this possible explanation, we have computed the band structure and spin polarizations for a slab of 5 MoTe$_2$ trilayers. In Figures S26a-c, we present the different components of the spin polarizations taken from the topmost trilayer of the slab. The presence of a mirror which reverts $k_y \to -k_y$ and $s_x \to -s_x$, $s_z \to -s_z$ can be seen explicitly. Using the Edelstein effect as a possible explanation, we now consider what happens when an electric field is applied to the sample which biases the population of the Fermi surface in the direction of the applied field. While the total spin of equilibrium Fermi surface must average to zero, the biasing can induce an average spin magnetization. For example, biasing the population in the $y$-direction can generate a finite average spin of $s_x$, as for the usual Rashba surface states. However, it is also clear that a bias in the $y$-direction can never generate a non-zero total $s_y$ because of the mirror symmetry. Considering now the case with 1% (Figures 26d-f) and 5% (Figures 26g-i) shear strain, we see that spin polarizations do not show any symmetry beyond time-reversal, and a bias in the $y$-direction can generate finite averages of all components.

We employed density functional theory (DFT) as implemented in the Vienna Ab Initio Simulation Package (VASP)[11,12]. For the VASP calculations, the exchange correlation term is described according to the Perdew-Burke-Ernzerhof (PBE) prescription together with projected augmented-wave pseudopotentials[13]. We calculated the surface energy cuts by using a slab geometry along the (001) direction for 5 MoTe$_2$ trilayers, achieving a negligible interaction between the surface states from both sides of the slab and reduce the overlap between top and bottom surface states, we considered a slab of 10-unit cells and 1 nm vacuum thickness and the kinetic energy cut off was set to 400 eV. For the energy cuts, we used a 20x20 grid of K points. Spin resolved figures were done using PyProcar package[14].

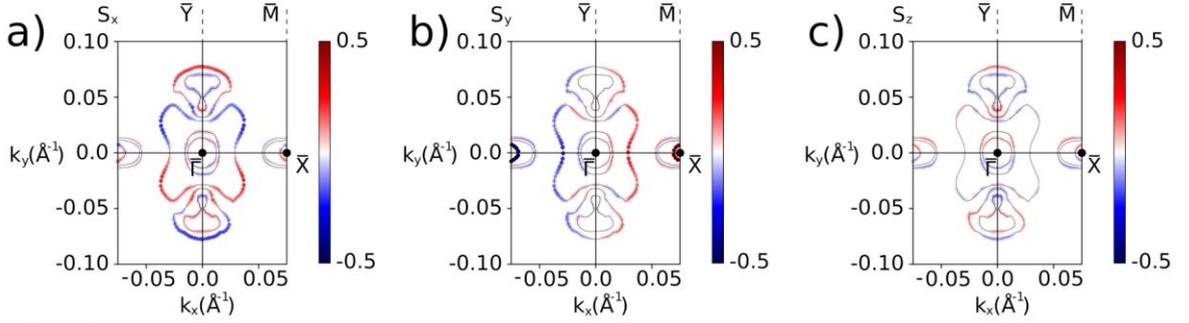

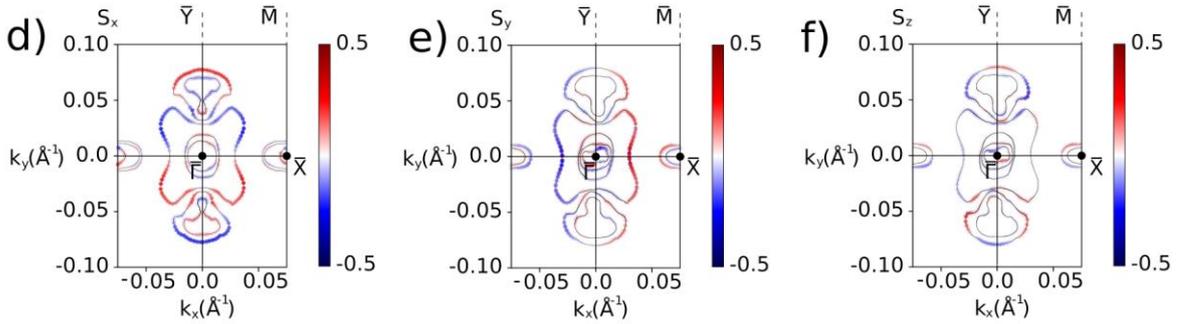

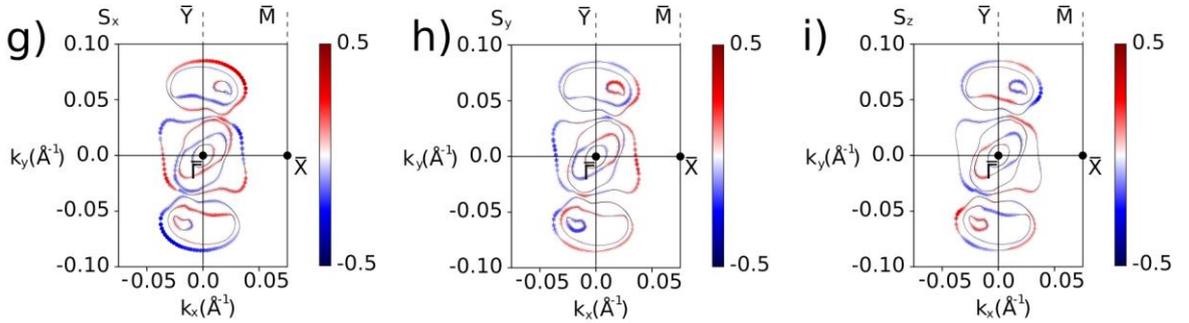

**Figure S26.** Band structure at the $k_x - k_y$ plane (in $\pi/a$) where the spin-texture is calculated for each spin component along the cartesian axis (*i.e.*, $s_x$, $s_y$ and $s_z$) at the Fermi level. (a-c) Band structure for 0% strain, (d-f) for 1% strain, and (g-i) for 5% strain. Figures were created by using the script PyProcar[14].

Symmetry constraints also apply to the bulk spin Hall effect and define the permitted spin to charge interconversion configurations, which appear as non-zero components in the spin Hall conductivity tensor. For simplicity, we describe here the symmetry constraints for the direct spin Hall conductivity tensor $\sigma_{ij}^{k}$, which connects a spin current $j_i^k$ (index $i$ indicates the spin diffusion direction and $k$ the spin polarization orientation) with an electric field $E_j$ generated along $j$ direction:

$$j_i^k = \sigma_{ij}^k E_j$$

Under time reversal symmetry, the spin Hall conductivity tensor for the space group P2$_1$/m (#11) to which the 1T' phase of MoTe$_2$ belongs has the following form:

$$\sigma_{ij}^{x} = \begin{pmatrix} 0 & \sigma_{xy}^{x} & 0 \\ \sigma_{yx}^{x} & 0 & \sigma_{yz}^{x} \\ 0 & \boldsymbol{\sigma_{zy}^{x}} & 0 \end{pmatrix}, \quad \sigma_{ij}^{y} = \begin{pmatrix} 0 & 0 & \sigma_{xz}^{y} \\ 0 & 0 & 0 \\ \sigma_{zx}^{y} & \boldsymbol{0} & 0 \end{pmatrix}, \quad \sigma_{ij}^{z} = \begin{pmatrix} 0 & \sigma_{xy}^{z} & 0 \\ \sigma_{yx}^{z} & 0 & \sigma_{yz}^{z} \\ 0 & \boldsymbol{\sigma_{zy}^{z}} & 0 \end{pmatrix},$$

where $\sigma_{ij}^{k} = -\sigma_{ji}^{k}$, which has been adapted from the Tensor utility at the Bilbao Crystallographic server[15]. The bold case highlights the components that can be detected in our experimental configuration.

For a high-symmetry crystal, for example any phase with two orthogonal mirror symmetries, only the mutually orthogonal components would be allowed. Due to the low symmetry of the 1T' phase, we observe that some non-orthogonal components can be different to zero.

In our experimental geometry, we measure the electric field associated to the charge current $\boldsymbol{j_c}$ generated along the $y$-axis. Since we are considering a bulk effect, spin current must be injected from the graphene into the MoTe$_2$ flake following the $z$-direction. With those constraints, the only components of the spin Hall conductivity tensor relevant to our experiments are the ones marked in bold in the tensor. One of them is just the conventional component $\sigma_{zy}^{x}$, which accounts for charge conversion of spins polarized along the $x$-axis. The other non-zero component is $\sigma_{zy}^{z}$, that leads to spin-to-charge conversion for spins injected with polarization along $z$. As discussed in the main text, even if allowed, this component is expected to be too small to be observed in our experiments. In contrast, the unconventional component that we experimentally observe ($\sigma_{zy}^{y}$) is not permitted by the mirror $\mathcal{M}_y$ symmetry of the 1T' bulk. In analogy to the discussion presented above for the Edelstein effect, only the breaking of the mirror symmetry $\mathcal{M}_y$ in our sample could account for $\sigma_{zy}^{y} \neq 0$ that is observed in our experiment.